\documentclass[aps,reprint,twocolumn,showpacs,preprintnumbers,amsmath,amssymb,nofootinbib,superscriptaddress,showkeys]{revtex4}

\usepackage{epsfig}
\usepackage{color}
\usepackage[T1]{fontenc}
\usepackage{txfonts}

\begin{document}

\title{Power Counting in Peripheral Partial Waves: The Singlet Channels}
  \author{M. Pav\'on Valderrama}\email{pavonvalderrama@ipno.in2p3.fr}
  \affiliation{School of Physics and Nuclear Energy Engineering, Beihang University, Beijing 100191, China}
  \affiliation{Institut de Physique Nucl\'eaire, CNRS-IN2P3, Univ. Paris-Sud, Universit\'e Paris-Saclay, F-91406 Orsay Cedex, France}
  \author{M. S\'anchez S\'anchez}\email{sanchezmario@ipno.in2p3.fr}
  \affiliation{Institut de Physique Nucl\'eaire, CNRS-IN2P3, Univ. Paris-Sud, Universit\'e Paris-Saclay, F-91406 Orsay Cedex, France}
  \author{C.-J. Yang}\email{yangjerry@ipno.in2p3.fr}
  \affiliation{Institut de Physique Nucl\'eaire, CNRS-IN2P3, Univ. Paris-Sud, Universit\'e Paris-Saclay, F-91406 Orsay Cedex, France}
  \author{Bingwei Long}\email{bingwei@scu.edu.cn}
  \affiliation{Department of Physics, Sichuan University, 29 Wang-Jiang Road, Chengdu, Sichuan 610064, China}
  \author{J. Carbonell}\email{carbonell@ipno.in2p3.fr}
  \affiliation{Institut de Physique Nucl\'eaire, CNRS-IN2P3, Univ. Paris-Sud, Universit\'e Paris-Saclay, F-91406 Orsay Cedex, France}
  \author{U. van Kolck}\email{vankolck@ipno.in2p3.fr}
  \affiliation{Institut de Physique Nucl\'eaire, CNRS-IN2P3, Univ. Paris-Sud, Universit\'e Paris-Saclay, F-91406 Orsay Cedex, France}
  \affiliation{Department of Physics, University of Arizona, Tucson, AZ 85721, USA}
\date{\today}

\begin{abstract} 
\rule{0ex}{3ex} 
We analyze the power counting of the peripheral singlet partial waves
in nucleon-nucleon scattering.
In agreement with conventional wisdom, we find that pion exchanges
are perturbative in the peripheral singlets. 
We quantify from the effective field theory perspective
the well-known suppression induced by the centrifugal barrier in the pion-exchange interactions.
By exploring perturbation theory up to fourth order, 
we find that the one-pion-exchange potential in these channels is demoted from leading to
subleading order by a given power of the expansion parameter that
grows with the orbital angular momentum.
We discuss the implications of these demotions for few-body calculations:
though higher partial waves have been known for a long time to be irrelevant 
in these calculations (and are hence ignored), here we explain
how to systematize the procedure in a way that is compatible
with the effective field theory expansion.
\end{abstract}


\maketitle

The derivation of nuclear forces from first principles is one of
the central problems of nuclear physics.
In fact the nuclear potential -- though not an observable -- is
the theoretical object from which we usually derive the properties
of nuclei and calculate their reaction rates.
With the recent onset of {\it ab initio} calculations, the direct link
between nuclear forces and nuclear structure can be established
for the first time, at least for light nuclei.
Yet the understanding of nuclear forces from quantum chromodynamics (QCD)
remains incomplete.

Nowadays, among the strategies to explain nuclear observables from QCD
the most promising are lattice QCD and effective field theory (EFT).
The first approach tries to calculate nuclear properties directly from QCD.
Lattice QCD, though close to, is still not able to explain light nuclei
in the physical world, i.e. for the physical pion mass.
First calculations for pion masses larger than the physical one
are promising~\cite{Beane:2010em}
and it is apparent that $m_{\pi} = 140\,{\rm MeV}$ will be reached soon.
The second approach -- the EFT formulation -- establishes the connection
between nuclear physics and QCD indirectly, avoiding the requirement of
complex lattice QCD calculations
(see Refs.~\cite{Bedaque:2002mn,Machleidt:2011zz} for reviews).
Nuclear EFT is the renormalization group evolution of QCD 
at low energies and thus equivalent to QCD.
The idea is to write down the most general Lagrangian involving
the low-energy fields and respecting the low-energy symmetries
of QCD, in particular chiral symmetry.
In this regard nuclear EFT simply amounts to rewriting the QCD Lagrangian
in terms of a different set of fields, nucleons and pions
instead of quarks and gluons.

A key ingredient of nuclear EFT is the existence of a separation of scales.
Pions and nucleons are no longer valid degrees of freedoms at momenta
above a characteristic hard scale $M$ of the order of $0.5-1.0\,{\rm GeV}$.
Yet most nuclear processes of interest happen at a softer scale $Q$
of the order of $100-200\,{\rm MeV}$, which can be identified
with the typical momenta of nucleons inside nuclei
or with the pion mass.
Moreover on a first approximation naive dimensional analysis (NDA)
suggests that the dimensionful couplings of the EFT might be natural
when expressed in units of the hard scale $M$.
From this point of view $M$ can be interpreted as the natural
QCD mass scale for hadrons $M \sim M_{\rm QCD}$ (of the order of
the rho or the nucleon mass, not to be confused with 
the QCD perturbative scale $\Lambda_{\rm QCD}$)
or $4 \pi f_{\pi}$.
As a consequence we can write the scattering amplitude in EFT
as a power series in terms of the ratio between the soft-scale $Q$
and the hard-scale $M$:
\begin{eqnarray}
T = \sum_{\nu} T^{(\nu)} {\left( \frac{Q}{M} \right)}^{\nu} \, .
\end{eqnarray}
This property goes under the name of power counting and
is probably the most useful feature of EFT.
Power counting implies that we can organize the calculations to achieve
a given goal of accuracy and lies at the core of the systematics
of the EFT approach.

We stress that the existence of a power counting depends
on {\it renormalizability}.
EFT expresses the interactions among its fields as a low-energy expansion.
While at low momenta $Q$ each term in this expansion is smaller
than the previous one, at high momenta $M$ the contrary happens
and eventually the EFT interactions diverge in the ultraviolet.
This does not pose a conceptual problem because EFT is only expected
to work at low energies.
But it poses a technical one: the ultraviolet divergences threaten to destroy
the low-energy EFT expansion via loop corrections.
The way to deal with this issue is {\it renormalization}: we introduce
a cut-off $\Lambda$ to regularize the loops, include counterterms to
absorb the $\Lambda$-dependence and finally check that the low-energy
observables do not depend on the choice of $\Lambda$.
Cut-off independence, i.e. renormalizability, is therefore a necessary
condition to have power counting in EFT.
Renormalizability and power counting are intimately intertwined, a feature
that is emphasized for instance in Wilsonian
renormalization~\cite{Wilson:1973jj}.
If renormalizability is absent 
at a certain order we have to include new
contact interactions until cut-off independence is restored,
which in turn constrains the power counting.
If for some reason our prejudices about power counting do not lead
to a renormalizable EFT, they have to be revisited and amended.

Traditional EFT knowledge is perturbative, chiral perturbation theory (ChPT)
being a good example.
The understanding of the renormalization group analysis of non-perturbative
physics -- as is the case in nuclear physics -- is recent in comparison.
But the first formulation of nuclear EFT, the Weinberg
counting~\cite{Weinberg:1990rz,Weinberg:1991um}, predates these developments.
To deal with the non-perturbative features of nuclear physics,
the Weinberg counting applies
not directly to the scattering amplitude but rather 
to the nuclear potential:
\begin{eqnarray}
V_{\rm EFT} = V^{(0)} + V^{(1)} + V^{(2)} + \dots \, .  
\end{eqnarray}
The rationale is that, though power counting is not obvious for the scattering
amplitude in the non-perturbative case, it is for the potential.
Thus the idea is to iterate the EFT potential, or at least the leading order
(${\rm LO}$) piece, in the Lippmann-Schwinger (or Schr\"odinger) equation with the implicit
expectation that the resulting $T$-matrix follows the EFT expansion
\begin{eqnarray}
T_{\rm EFT} = T^{(0)} + T^{(1)} + T^{(2)} + \dots \, .  
\end{eqnarray}
Taking into account the simplicity of the approach outlined by Weinberg
it comes as no surprise the enthusiasm with which it has been followed,
leading to the development of two-nucleon potentials with
$\chi^2/\rm{d.o.f.} \sim 1$~\cite{Entem:2002sf,Entem:2003ft,Epelbaum:2004fk}.

Yet there is a fly in the ointment: it is not clear whether the scattering
amplitudes follow a power counting in Weinberg's proposal.
In fact the situation is rather the contrary.
Several works have shown once and again~\cite{Kaplan:1996xu,Beane:2001bc,Nogga:2005hy,Valderrama:2005wv,PavonValderrama:2005uj,Entem:2007jg,Yang:2009kx,Yang:2009pn,Yang:2009fm,Zeoli:2012bi,Marji:2013uia} 
that the Weinberg counting is not renormalizable.
As the cut-off is increased above $\Lambda \sim 0.5\,{\rm GeV}$,
the scattering amplitudes show all kinds of fascinating behavior
with the notable exception of cut-off independence.
The most natural interpretation of the previous results is that renormalization
is lost in the Weinberg counting.
Not only that, the connection with QCD disappears
because this 
approach does not make for a proper EFT.
We mention in passing that a few authors have proposed to reinterpret
what renormalization means in order to accommodate the previous observations
in the Weinberg counting~\cite{Epelbaum:2006pt,Epelbaum:2009sd}.
This is adequate as long as one does not insist on a clear
power counting (beyond the vague impression that calculations get better
at higher orders) and a direct connection to QCD.
However this is not the solution for an appropriate nuclear EFT.
Rather we should strive to get scattering amplitudes that are unambiguously
renormalizable and have a power counting.

As a matter of fact there exists already a consistent version of nuclear EFT~\cite{Nogga:2005hy,Birse:2005um,Valderrama:2009ei,Valderrama:2011mv,Long:2011qx,Long:2011xw,Long:2012ve}.
It is renormalizable, can describe the scattering amplitudes for $k < M$,
converges well and power counting is realized at the level of observables.
Its foundation relies on a better understanding of the renormalization of
non-perturbative physics and singular interactions~\cite{Beane:2000wh,Beane:2001bc,Nogga:2005hy,Valderrama:2005wv,PavonValderrama:2005uj,Long:2007vp}.
The key improvements over the original Weinberg proposal are
the non-perturbative renormalization of the ${\rm LO}$ amplitudes and
the addition of subleading order contributions as perturbations.
At ${\rm LO}$ the main difference with the Weinberg counting lies
in the promotion of a series of $P$- and $D$-wave counterterms to ${\rm LO}$
in triplet partial waves for which the tensor force is attractive,
a change 
proposed in Ref.~\cite{Nogga:2005hy}.
At subleading orders there are more counterterms than in Weinberg counting,
for instance in the attractive triplets that already received
a counterterm at ${\rm LO}$.
The convergence of the EFT expansion is acceptable and the description of
the data too, but it has not achieved yet a $\chi^2 / {\rm d.o.f.} \sim 1$
as in the Weinberg approach.
However this is expected if we take into account that the calculations of
Refs.~\cite{Valderrama:2009ei,Valderrama:2011mv,Long:2011qx,Long:2011xw,Long:2012ve}
are still one order below the most advanced ones in the Weinberg
approach~\cite{Entem:2002sf,Entem:2003ft,Epelbaum:2004fk}.

One prediction of the original Weinberg counting is that the one-pion-exchange (OPE) potential
-- the ${\rm LO}$ nuclear potential -- is {\it always} non-perturbative.
Yet this is in contradiction with common wisdom in nuclear physics.
Pion exchanges have been known for a long time to be perturbative
in peripheral partial waves~\cite{Kaiser:1997mw,Kaiser:1998wa}.
While this is easy to understand in terms of the repulsive
centrifugal barrier for high angular momenta ($l \gg k / m_{\pi}$),
the analogous explanation in terms of power counting
has traditionally remained elusive.
From time to time this has been discussed in the literature~\cite{Nogga:2005hy,Birse:2005um,Valderrama:2011mv,Long:2011xw},
but rather as an afterthought.
This peripheral demotion already happens for moderate angular momenta
$l \sim k / m_{\pi}$. Different arbitrary angular momentum cut-offs
for when to consider OPE perturbative (or not) have been proposed,
usually in the range $l = 2-4$.
However, taking into account that the importance of this matter is rather
tangential in two-body calculations, it comes as no surprise
the disinterest for this particular problem~\footnote{
Actually power counting takes into account the peripheral suppression
in the particular case of the contact-range interactions, where
the $l$-wave counterterms are less relevant than the $S$-wave ones
by a factor of $Q^{2l}$ in naive dimensional analysis.}.
The task of power counting should be to quantify the size of the peripheral
wave suppression to systematically include it in EFT calculations.

The answer to this question is not only of academic interest,
but has practical implications too.
The main applications are few-body calculations, which usually require
the inclusion of contributions arising from two-body partial waves
up to a critical value of the orbital angular momentum
(typically $l \geq 5$ or $j \geq 5$ in the three-nucleon
system~\cite{Witala:2000am}).
Yet the choice of a maximum angular momentum is driven by numerical
considerations rather than by the constraints that the EFT expansion
imposes on the accuracy of physical observables.

The purpose of this manuscript is to translate the quantum-mechanical
peripheral-wave suppression into a power-counting demotion of
the OPE potential.
This will allow to take the decisions regarding which two-body partial waves
to ignore based on power-counting arguments.
That means improving the systematics of these calculations
or even simplifying them at the lowest orders where probably
very few partial waves need to be included.
Here we limit ourselves to the spin-singlet waves
where OPE is not singular and thus can be defined
without counterterms.

The manuscript is structured as follows.
In Section \ref{sec:perturbative}
we will compare the non-perturbative OPE predictions for the phase
shifts in the singlets with their perturbative expansion.
This will allow us to check up to what point OPE is perturbative.
In Section \ref{sec:demotion} we will find a power-counting explanation
for the peripheral demotion of central OPE,
which will
later be checked against numerical calculations of the expansion
parameter of central OPE.
Finally we will present our conclusions in Section \ref{sec:conclusion}.

\section{Perturbative Pions in Peripheral Singlet Waves}
\label{sec:perturbative}

In this section we will analyze whether the OPE potential is perturbative
in the peripheral singlet waves, that is, the $^1P_1$, $^1D_2$, $^1F_3$, $\dots$
partial waves.
The approach is straightforward: we will compare the amplitudes
that we obtain after the full iteration of OPE with
the perturbative ones.
We will work out the phase shifts up to fourth order in perturbation theory.
The outcome of these lengthy calculations is that the OPE potential
is definitely perturbative in all the singlet waves with orbital angular
momentum $l \geq 1$ (the $l = 0$ case will be dealt with
in a 
separate manuscript).
The interpretation in terms of traditional power counting is
that the OPE potential is 
subleading (where ${\rm LO}$ is reserved for
interactions that have to be iterated to reproduce the non-perturbative
physics of the $S$-waves).

\subsection{Formalism}

Here we will determine the perturbative character of OPE
in the peripheral singlets by means of a direct calculation.
In first place we will solve the Schr\"odinger equation
with the OPE potential to obtain the non-perturbative
phase shift $\delta_{l}$, where $l$ indicates
the orbital angular momentum.
In a second step we will expand the phase shift $\delta_{l}$ perturbatively as
\begin{eqnarray}
\delta_l = \delta_l^{[1]} + \delta_l^{[2]} + \delta_l^{[3]} + \dots \, ,
\end{eqnarray}
where the numbers in brackets indicate the number of insertions
of the OPE potential.
Comparing the right- and left-hand sides of the equation above,
we can check the convergence of the perturbative series.

At this point we find it worth commenting that this kind of direct comparison
can only be used in the singlets.
As it is well-known, the OPE potential comprises central and tensor pieces.
The former is regular, and hence the phase shifts and
their perturbative expansion are well-defined.
But the latter is singular: to determine whether the tensor piece is
perturbative or not requires renormalization.
In the singlet channels only the central piece of the OPE potential
is involved in the calculations.
This is what makes possible the simple analysis we pursue in this section.

The calculations are done as follows.
For the phase shifts, we begin with the reduced Schr\"odinger equation:
\begin{eqnarray}
- {u_l}'' + \left[ 2 \mu V(r) + \frac{l(l+1)}{r^2}
\right]\,u_l(r) = k^2\,u_l(r) \, ,
\end{eqnarray}
where $V$ is the OPE potential.
We solve it with a regular boundary condition at the origin,
i.e. $u(0) = 0$.
We can extract the phase shifts from the asymptotic behavior of
the wave function at large distances ($m_{\pi} r \gg  1$):
\begin{eqnarray}
u_l(r; k) \to \hat{\jmath}_l(k r) - \tan{\delta_l} \hat{y}_l(k r) \, , 
\label{eq:asymptotic}
\end{eqnarray}
where $\hat{\jmath}_l(x) = x\,j_l(x)$, $\hat{y}_l(x) = x\,y_l(x)$ are reduced
spherical Bessel functions.
For obtaining the non-perturbative phase shifts, we will match
the asymptotic expression above to the reduced wave function
$u_l$ at the infrared cut-off $R = 20\,{\rm fm}$.

In contrast, if we are interested in the perturbative phase shifts,
the asymptotic matching will not be the method of choice.
We will find much more useful instead the following integral representation
of the phase shift:
\begin{eqnarray}
\tan{\delta}_l = -\frac{2\mu}{k}\,\int_{0}^{\infty}\,u_k(r)\,V(r)\,
\hat{\jmath}_l(k r)\,dr ,
\label{eq:integral}
\end{eqnarray}
in which the wave function $u_k$ is normalized as in Eq.~(\ref{eq:asymptotic}).
The advantage of the integral formula above is that it is better suited for
a perturbative calculation: with a wave function of order $n$ 
we can calculate the phase shift at order $n+1$.
A derivation of this formula can be found in Appendix \ref{app:integral}.

If the potential is weak enough, we can expand the reduced wave function
perturbatively as follows
\begin{eqnarray}
u_l(r; k) = u_l^{[0]} + u_l^{[1]} + u_l^{[2]} + u_l^{[3]} + \dots \, , 
\end{eqnarray}
where the brackets indicate the number of OPE insertions.
The differential equations followed by the $u_l^{[n]}$'s are then
\begin{eqnarray}
{u_l^{[0]}}'' + \left[ k^2 - \frac{l(l+1)}{r^2}\right]\,u_l^{[0]}(r) &=&
0 \, , \label{eq:pert0} \\
{u_l^{[n]}}'' + \left[ k^2 - \frac{l(l+1)}{r^2}\right]\,u_l^{[n]}(r) &=&
2 \mu V(r) \, u_l^{[n-1]}(r) \, 
\nonumber \\ \mbox{for $n \geq 1$,} \label{eq:pertn} && 
\end{eqnarray}
which we obtain from plugging the perturbative expansion of the wave function
into the reduced Schr\"odinger equation and rearranging the terms.

We solve the set of differential equations iteratively, starting with $n = 0$
for which we take $u^{[0]}_l(r; k) =\hat{\jmath}_l(k r)$, i.e. the free solution.
If we now expand the integral expression for the phase shifts
(Eq.~(\ref{eq:integral})) perturbatively,
we end up with the Born approximation
\begin{eqnarray}
\delta_l^{[1]} &=&
-\frac{2\mu}{k}\,\int_{0}^{\infty} u_l^{[0]}(k; r)\,
V(r)\,\hat{\jmath}_l(k r)\,dr  \, .
\end{eqnarray}
For $n \geq 1$ the only subtlety is finding a suitable boundary condition
for $u^{[n]}_l$.
Actually, the easiest thing to do is to integrate from infinity to the origin.
The reason is that the asymptotic form of $u^{[n]}_l(k; r)$ is readily available
from expanding Eq.~(\ref{eq:asymptotic}) perturbatively.
At first order we find the asymptotic boundary condition
\begin{eqnarray}
u_l^{[1]}(r; k) &\to& -\delta^{[1]}\,\hat{y}_l(k r) \, , 
\end{eqnarray}
from which we can integrate $u_l^{[1]}$ for arbitrary $r$.
Then we can calculate the second order contribution to the phase shifts
by inserting $u_l^{[1]}$ in the perturbative expansion of
Eq.~(\ref{eq:integral}), yielding
\begin{eqnarray}
\delta_l^{[2]} &=&
-\frac{2\mu}{k}\,\int_{0}^{\infty} u_l^{[1]}(r; k)\,
V(r)\,\hat{\jmath}_l(k r)\,dr  \, .
\end{eqnarray}
For the higher orders we simply repeat the previous steps.
The asymptotic boundary conditions and the expressions
for the $n=3,4$ contributions to the phase shifts read:
\begin{eqnarray}
u_l^{[2]} &\to& -\delta^{[2]}\,\hat{y}_l(k r) \, , \\ \nonumber \\
\delta_l^{[3]} &=&
-\frac{2\mu}{k}\,\int_{0}^{\infty}u_l^{[2]}\,V(r)\,\hat{\jmath}_l(k r)\,dr  
- \frac{1}{3}{\delta_l^{[1]}\,}^3 \, ,
\end{eqnarray}
\begin{eqnarray}
u_l^{[3]} &\to& -\left[ \delta^{[3]} + \frac{1}{3}{\delta^{[1]}\,}^3
\right]\,\hat{y}_l(k r) \, , \\ \nonumber \\
\delta_l^{[4]} &=&
-\frac{2\mu}{k}\,\int_{0}^{\infty}u_l^{[3]}\,V(r)\,\hat{\jmath}_l(k r)\,dr  
- {\delta_l^{[1]}\,}^2\,\delta_l^{[2]} \, .
\end{eqnarray}
Expressions for orders $n > 4$ can be obtained analogously.

\subsection{Perturbative OPE}

Now we evaluate the previous perturbative differential equations
and integrals to calculate the phase shifts.
We remind that the form of the central piece (the only one we need
in the singlets) of the OPE potential is given by
\begin{eqnarray}
V(r) = \frac{\sigma\,\tau}{3}\,\frac{1}{M_N \Lambda_{NN}}
\,m_{\pi}^3\,\frac{e^{-m_{\pi} r}}{m_{\pi} r} \, ,
\end{eqnarray}
where $\sigma = \vec{\sigma}_1 \cdot \vec{\sigma}_2$ ($ = -3$ for the singlets),
$\tau = \vec{\tau}_1 \cdot \vec{\tau}_2$, $M_N$ is the nucleon mass and
$\Lambda_{NN}$ is the characteristic momentum scale of OPE,
defined as
\begin{eqnarray}
\Lambda_{NN} = \frac{16 \pi f_{\pi}^2}{M_N\,g_A^2} \sim 300\,{\rm MeV} \, ,
\end{eqnarray}
with $g_A = 1.26$ and $f_{\pi} = 92.4\,{\rm MeV}$.
By {\it characteristic scale} we mean the following: we naively expect
the OPE to behave perturbatively for momenta $k < \Lambda_{NN}$,
while for higher momenta it becomes
non-perturbative~\cite{Kaplan:1998tg,Kaplan:1998we,Fleming:1999ee}.
As we will see, this is not the case in the peripheral singlets.

\begin{figure*}[htt!]
\begin{center}
\epsfig{figure=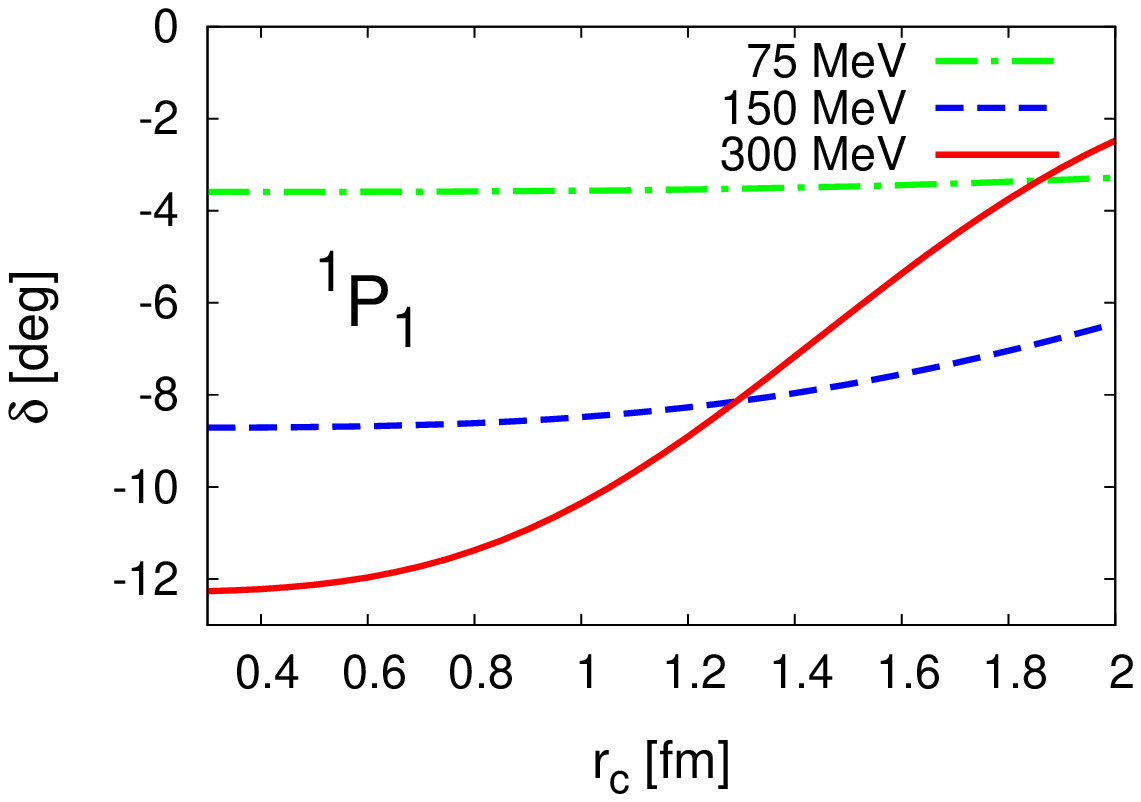,
	height=4.9cm, width=5.0cm}
\epsfig{figure=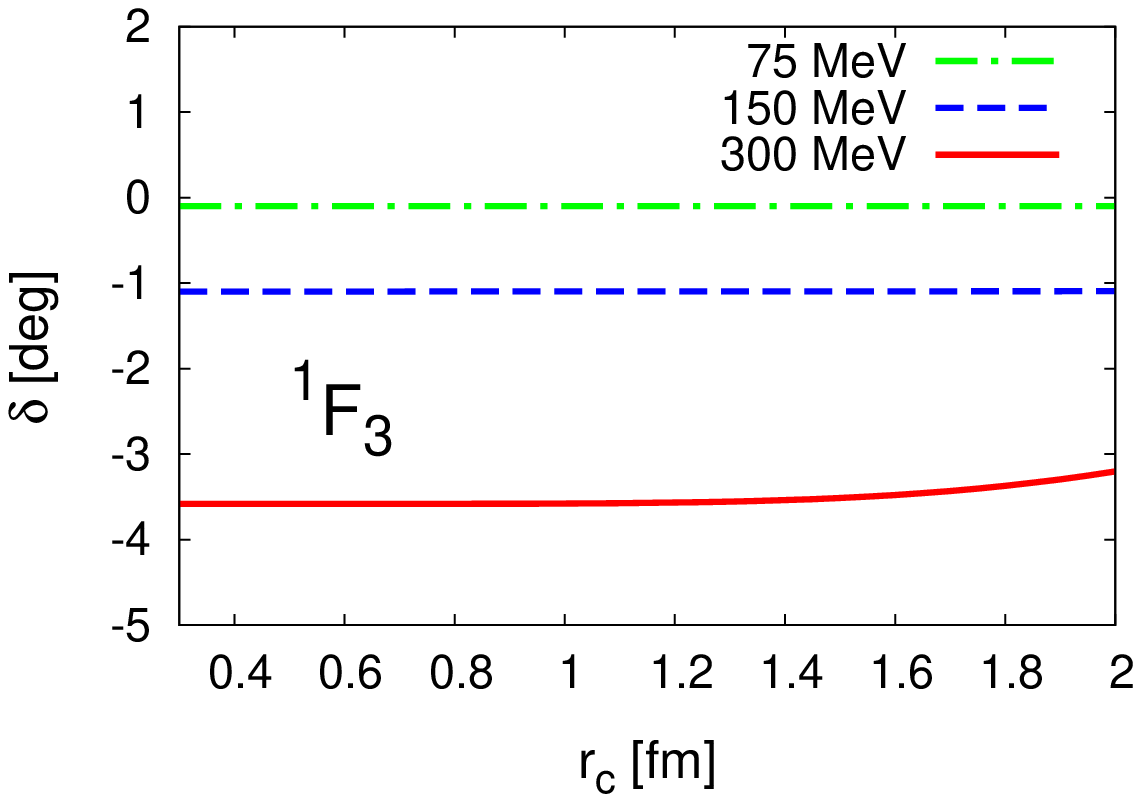,
	height=4.9cm, width=5.0cm}
\epsfig{figure=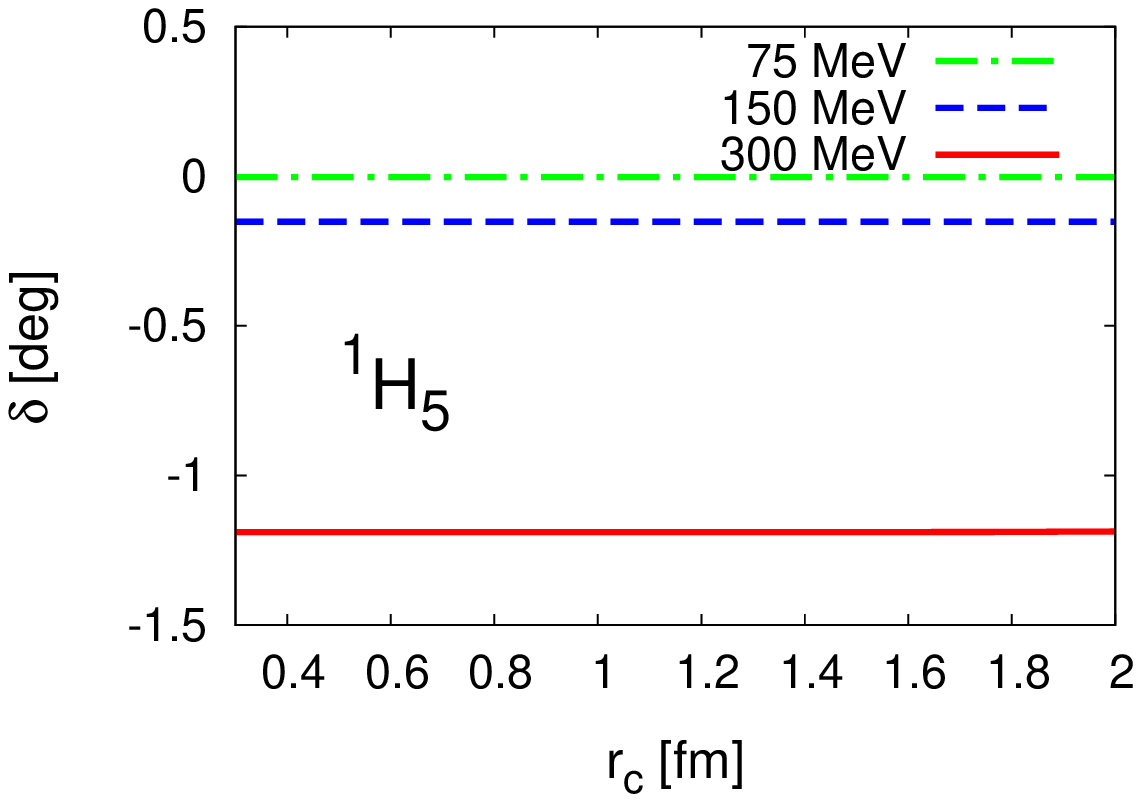,
	height=4.9cm, width=5.0cm}
\end{center}
\caption{Cut-off dependence of the non-perturbative phase shifts
for the OPE potential in the singlet peripheral waves
$^1P_1$, $^1F_3$ and $^1H_5$.
We show the isoscalar partial waves because for them the OPE potential
is three times stronger than in the isovector partial waves. Hence
the cut-off dependence is also expected to be more evident
in the isoscalar channels.
We show the $r_c = 0.3-2.0\,{\rm fm}$ cut-off range for the c.m. momenta
$k_{\rm c.m.} = 75, 150$ and $300\,{\rm MeV}$.
}
\label{fig:W-cutoff}
\end{figure*}

We have previously commented that the perturbative expansion is finite
and well-defined in the singlets.
Yet we will keep a finite cut-off in the calculations.
In particular we regularize the OPE potential as follows
\begin{eqnarray}
V(r; r_c) = V(r)\,\theta(r - r_c) \, ,
\end{eqnarray}
which simply amounts to changing the lower limit of the perturbative integrals
from $r=0$ to $r=r_c$.
The reasons for having a cut-off are:
i) in actual EFT calculations (including counterterms and the two-pion
exchange potential, which we have ignored in this work)
a finite cut-off is a must,
ii) the computational requirements of doing perturbation theory
to high order decrease significantly with a finite cut-off~\footnote{We are
using a variable step integration method for the set of coupled differential
equations that define the perturbative expansion, Eqs.~(\ref{eq:pert0}) and
(\ref{eq:pertn}). Owing to the number of differential equations involved
it is increasingly difficult to solve it for small cut-offs when i) we
increase the perturbative order for which we integrate and ii)
we increase the angular momentum, particularly at low momenta.
The chosen cut-off is actually on the limit of what we can compute
at fourth order, yet it is more than enough for a nuclear EFT calculation.}.
We have chosen $r_c = 0.3\,{\rm fm}$ as a reference value for the cut-off.
For this cut-off the perturbative calculations with OPE have already converged
(as a matter of fact, there are only tiny differences in the results
for $r_c < 1\,{\rm fm}$).
Besides, $r_c = 0.3\,{\rm fm}$ is also significantly below the standard
cut-off ranges employed in previous EFT calculations
in $r$-space~\cite{Valderrama:2009ei,Valderrama:2011mv}.
For an $S$-wave the equivalence of the previous $r$-space cut-off
with a sharp $p$-space cut-off is $\Lambda = 1033\,{\rm MeV}$,
where we have used the relationship
$\Lambda\,r_c = \pi / 2$.
This relationship can be obtained by comparing the running of the $S$-wave
counterterm $C_0$ in a pionless theory for a delta-shell $r$-space
regulator $r_c$ and a sharp cut-off $p$-space regulator $\Lambda$
(see Ref.~\cite{Entem:2007jg} for details).
The extension of the previous relationship for peripheral partial waves
is trivial, yielding the numerical values $\Lambda = 1590\,{\rm MeV}$
for a $P$ wave, $\Lambda = 2127\,{\rm MeV}$ for a $D$ wave and
higher values for $l \geq 3$.
For checking purposes we explicitly show the cut-off dependence of the isoscalar
partial waves in Fig. \ref{fig:W-cutoff}, where it can be appreciated that the 
more peripheral the wave the weaker the cut-off dependence.
We have chosen to show the isoscalar channels because it is for these
channels that central OPE is strongest, yielding more cut-off
dependence.
Yet, with the exception of the $^1P_1$ channel for $r_c \geq 1\,{\rm fm}$
(already a relatively large cut-off), the cut-off dependence
ranges from rather mild ($^1F_3$) to negligible ($^1H_5$).

\begin{figure*}[htt!]
\begin{center}
\epsfig{figure=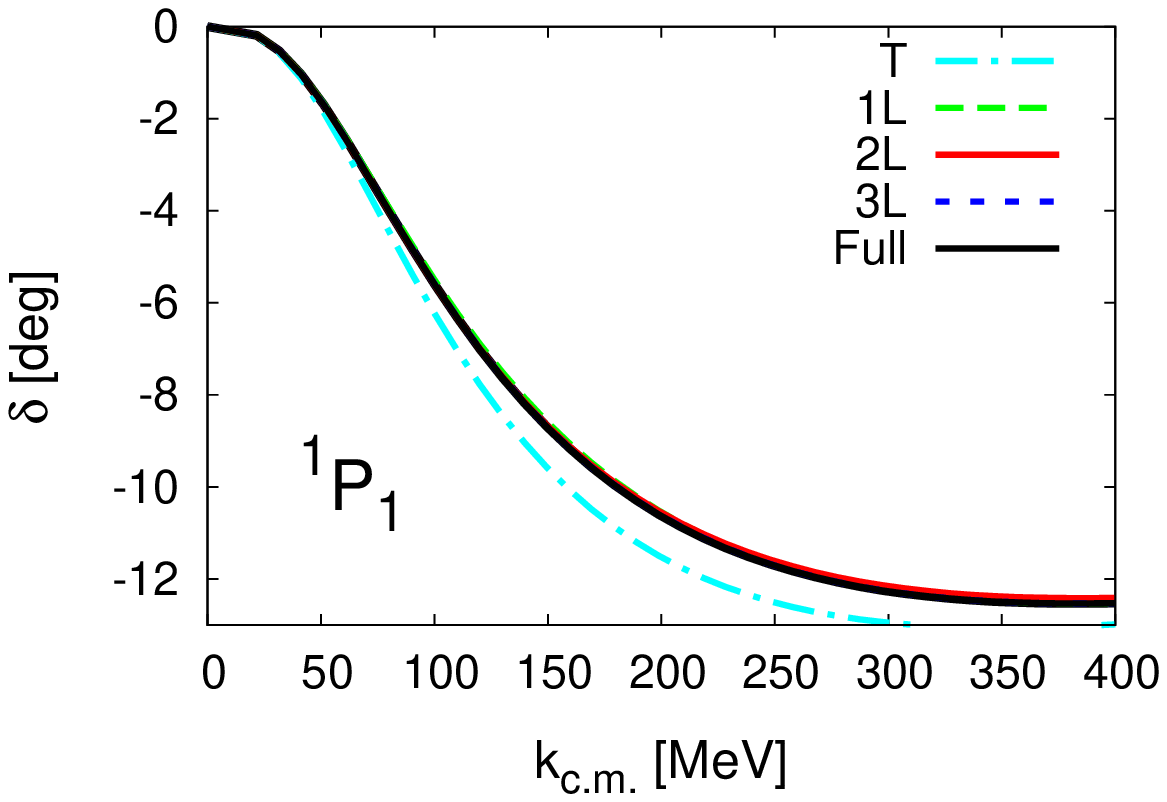,
	height=4.9cm, width=5.0cm}
\epsfig{figure=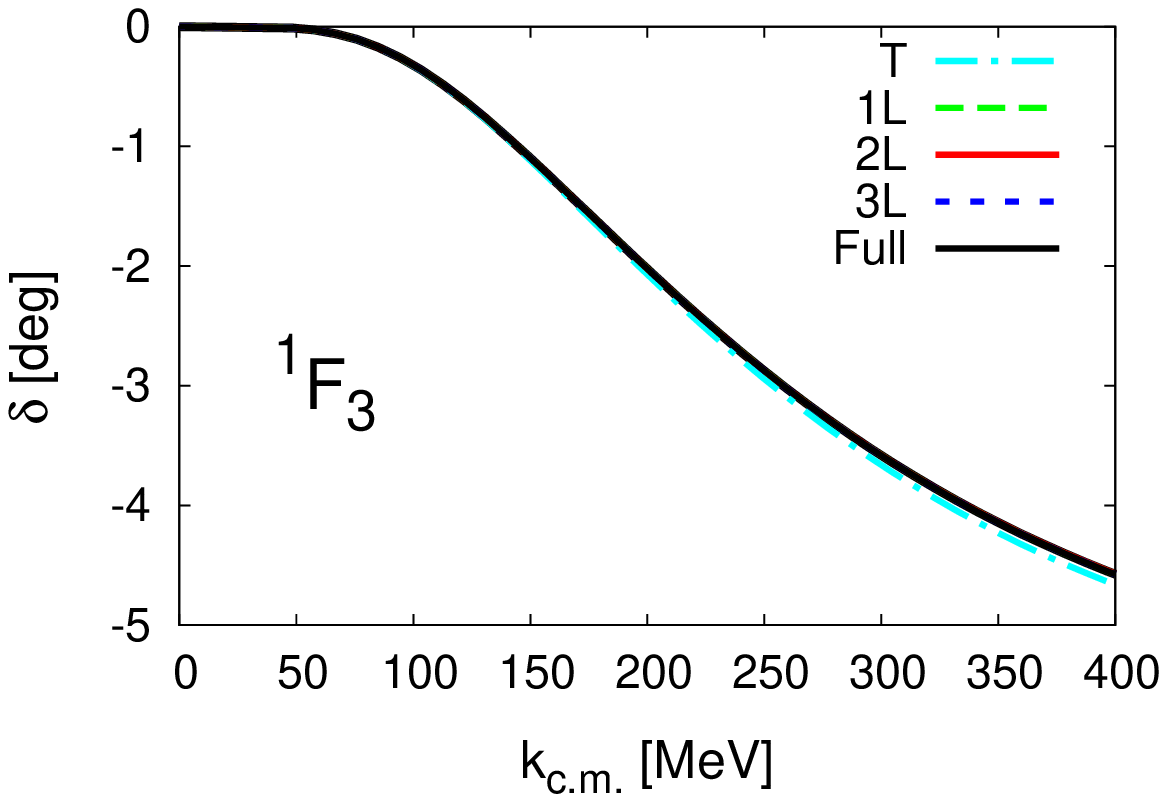,
	height=4.9cm, width=5.0cm}
\epsfig{figure=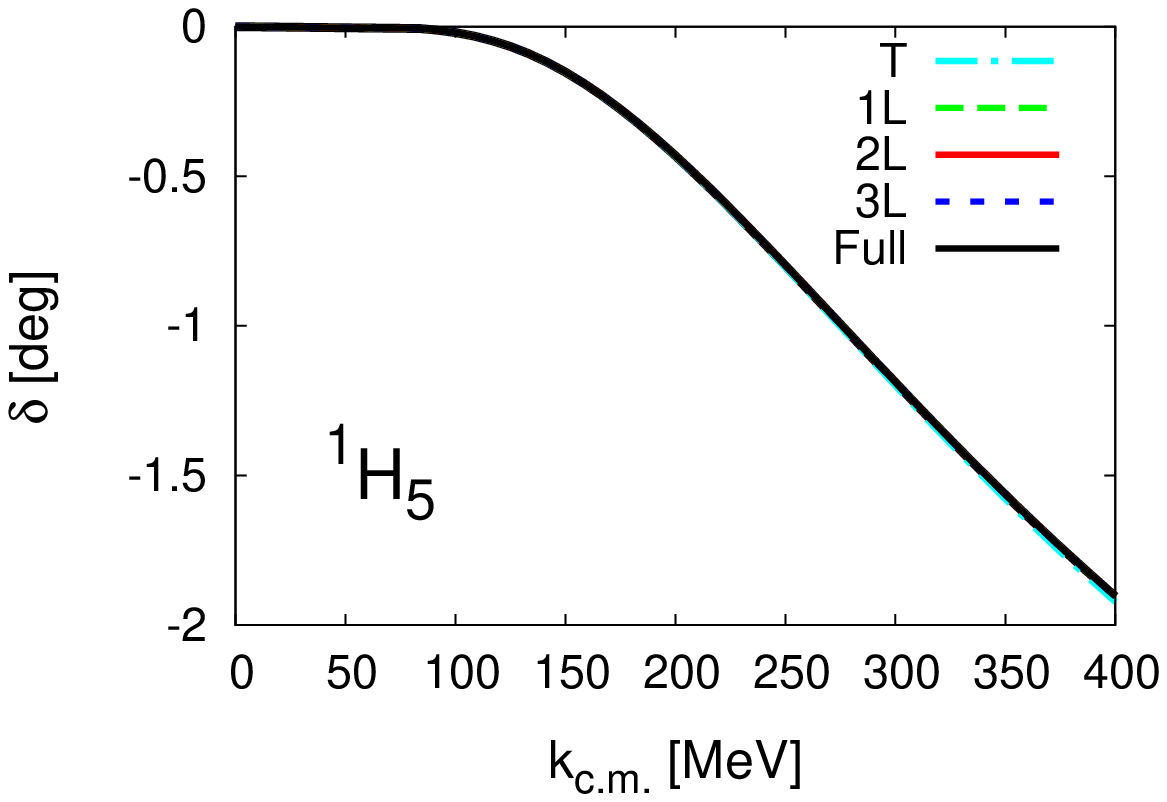,
	height=4.9cm, width=5.0cm}
\epsfig{figure=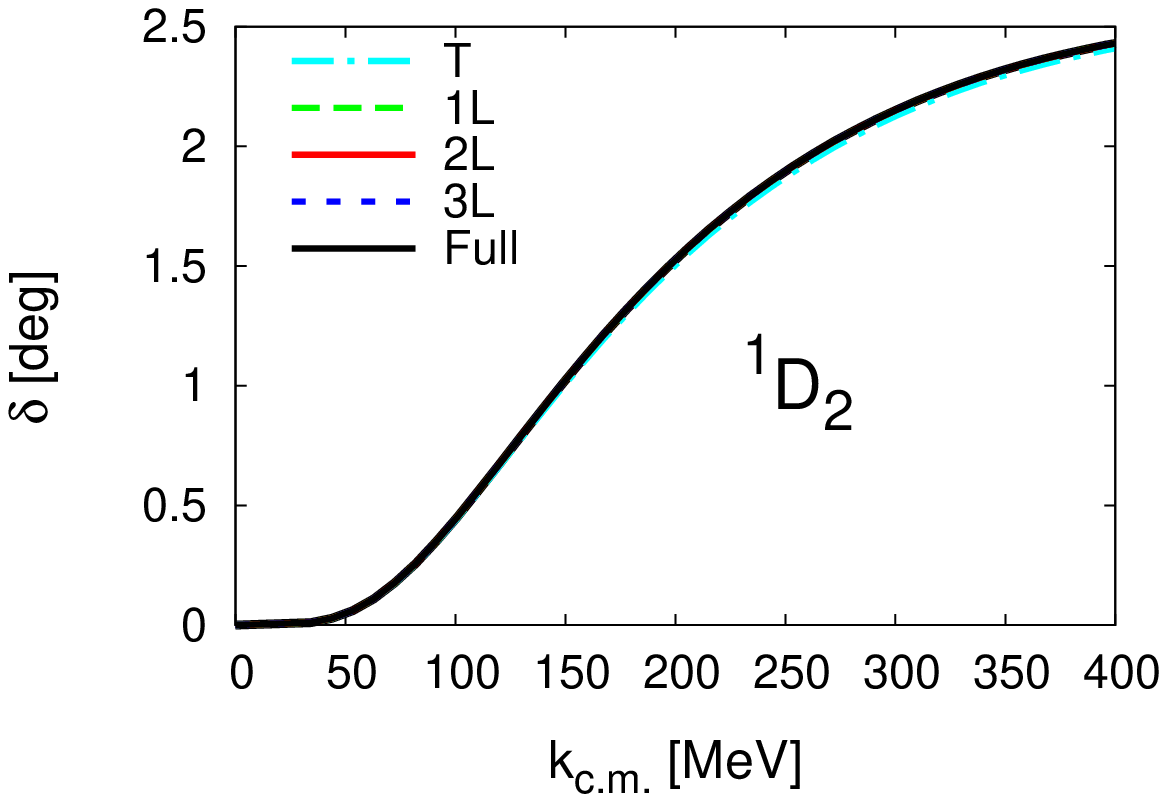,
	height=4.9cm, width=5.0cm}
\epsfig{figure=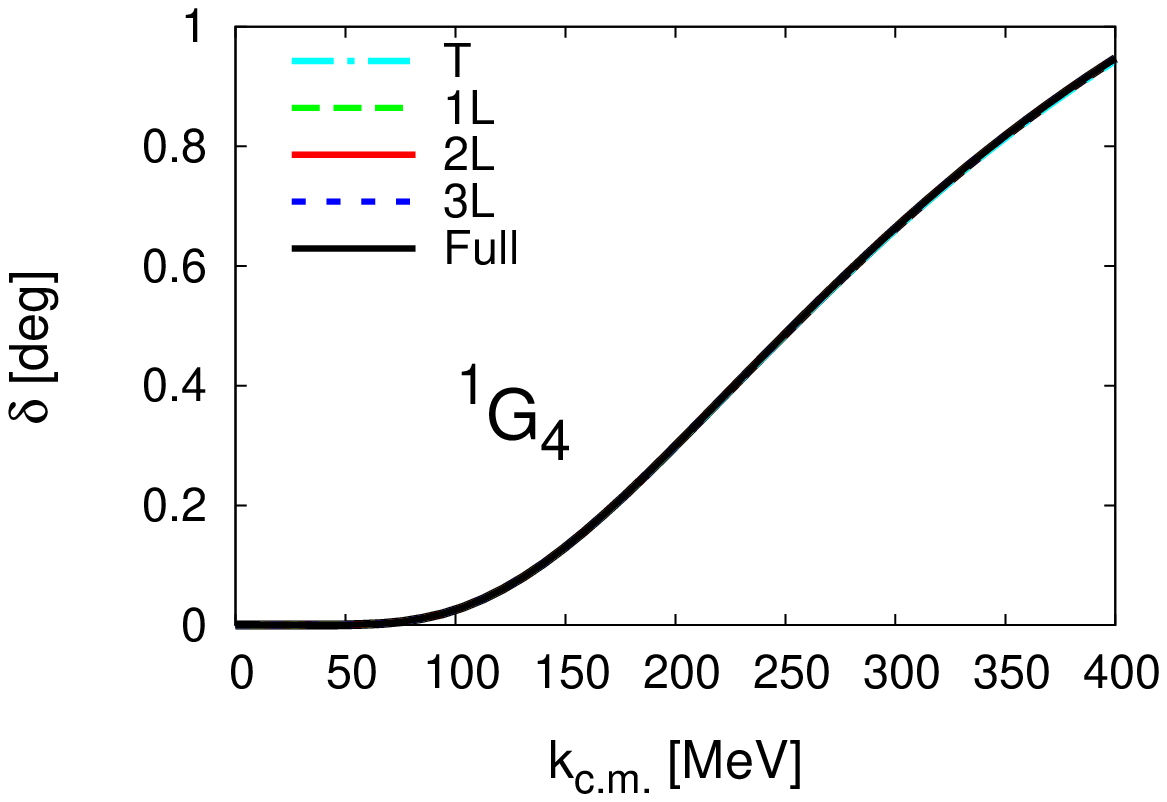,
	height=4.9cm, width=5.0cm}
\epsfig{figure=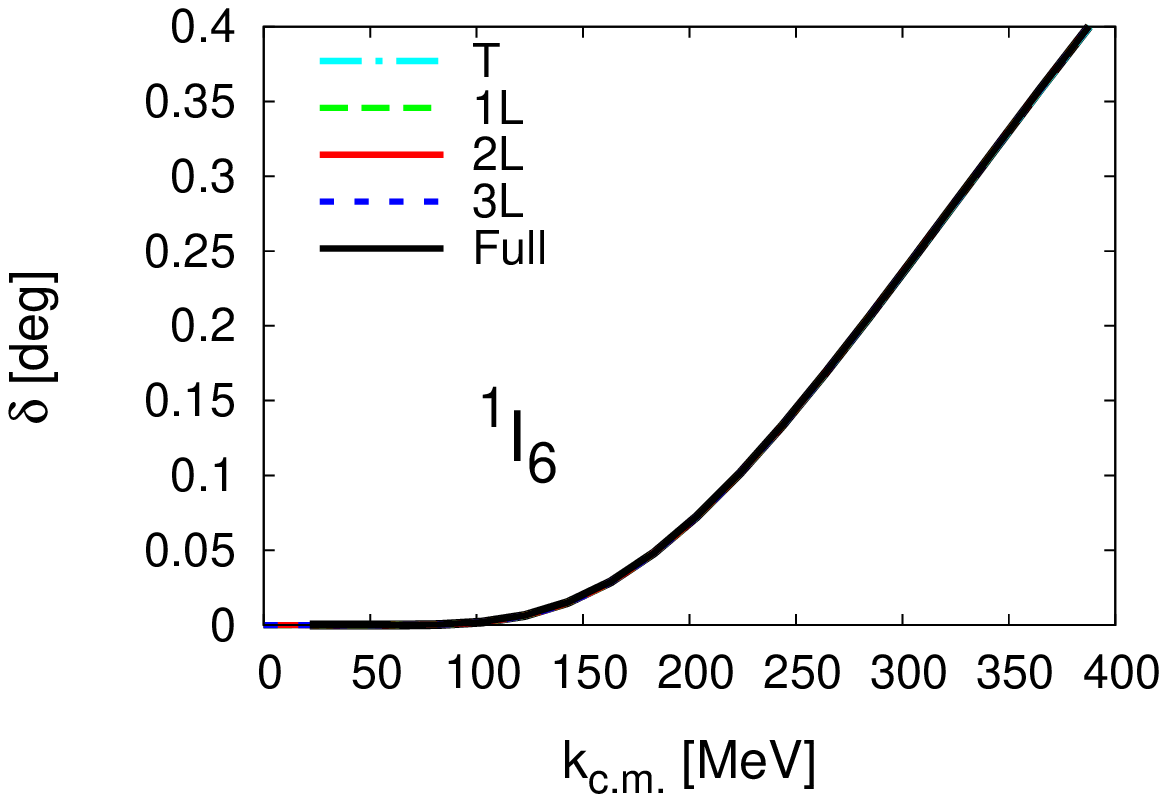,
	height=4.9cm, width=5.0cm}
\end{center}
\caption{Convergence of the perturbative expansion of the phase shifts
for the OPE potential in the singlet peripheral partial waves
$^1P_1$, $^1D_2$, $^1F_3$, $^1G_4$, $^1H_5$ and $^1I_6$.
We compare the full, non-perturbative OPE phase shifts -- the black solid line
in the figures -- with the perturbative ones computed at increasing orders,
where ``T'' stands for ``tree level'' (i.e. the Born approximation)
and ``1L'', ``2L'', ``3L'' for the one-, two- and three-loop calculation
(i.e. second, third and fourth order perturbation theory).
The calculations have been performed with a finite cut-off of
$r_c = 0.3\,{\rm fm}$.
}
\label{fig:W-OPE}
\end{figure*}

We obtain the phase shifts shown in Figure \ref{fig:W-OPE},
where we can appreciate that the perturbative expansion
is converging extremely quickly even for the $^1P_1$ wave,
the lowest partial wave considered.
The perturbative series is more convergent the higher the partial wave: the tree level (that is,
the Born approximation) phase shifts already match the non-perturbative
ones with a precision of a fraction of a degree, with the exception of the $^1P_1$ wave.
All this indicates that the convergence parameter of the perturbative
expansion is certainly smaller (i.e. a faster convergence) than that
of the EFT for the particular case of the singlet waves.

\section{The Peripheral Wave Demotion}
\label{sec:demotion}

In this section we will discuss the role of angular momentum
for the power counting of the singlets.
The aim is to explain the fast convergence rate of the perturbative
series for central OPE in the singlets.
As already seen, the iteration of the OPE potential is suppressed
in the peripheral waves with respect to the expectations
of standard power counting.
However, the problem is how to incorporate the peripheral suppression
into the EFT expansion.
For that we have to quantify the size of this effect.
We will discuss a series of ideas to find the factor
by which the iteration of the OPE potential
is suppressed in the higher partial waves.

\subsection{Quantum Mechanical Suppression}

We begin by considering the peripheral suppression of a finite-range potential
in standard quantum mechanics.
The arguments are straightforward and relatively well-known, but we find
it convenient to repeat them for illustrative purposes.
This type of suppression is only apparent at momenta far below
the inverse of the range of the potential, i.e. $k \ll m_{\pi}$
for nuclear forces.
This means that it will help to explain the power counting in pionless theories.
However, the application to pionful theories will be limited
to very peripheral waves.

\begin{figure}[ttt!]
\begin{center}
\epsfig{figure=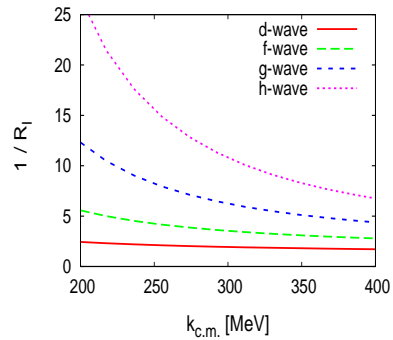,
	height=4.9cm, width=5.4cm}
\end{center}
\caption{Inverse of the ratio of the the diagonal potential $V_l$
in momentum space between the $l$-wave and the $p$-wave.
(see the main text for a detailed explanation of the ratio).
}
\label{fig:ratio-born}
\end{figure}

First we begin by considering the matrix element of a central potential
in the partial wave $l$, which is given by
\begin{eqnarray}
\langle p', l | V | p, l \rangle =
\frac{4\pi}{M_N}\,\int r^2 dr j_l(p' r) M_N V(r) j_{l} (p r) \, ,
\nonumber \\ \label{eq:V-central}
\end{eqnarray}
where $j_l(x)$ is the standard spherical Bessel function
and $V$ the potential in coordinate space.
In the Born approximation the size of this matrix element for $p = p'$
is proportional to the scattering amplitude.
If we compare
\begin{eqnarray}
V_l(p) = \langle p, l | V | p, l \rangle
\end{eqnarray}
for different angular momenta we can obtain a baseline estimation of
the peripheral suppression factor.
We can do this by calculating the ratio of $V_l$ against a reference
partial wave $l_0$.
A natural choice is the $P$ wave as it is the smallest angular momenta
considered in this work.
There is the complication that even (odd) partial waves
are isoscalars (isovectors).
This can be circumvented by taking into account the isospin factors
$\tau = \vec{\tau}_1 \cdot \vec{\tau}_2$ into the definition of
the ratio
\begin{eqnarray}
R_l(k) = \frac{\tau_{l_0}}{\tau_l}\,\frac{V_l(k)}{V_p(k)} \, ,
\end{eqnarray}
where $l_0 = 1$, $V_p=V_{^1P_1}$, and $\tau=1$ or $-3$ depending on the total isospin of a particular channel.
We have calculated the inverse of this ratio in Figure \ref{fig:ratio-born},
where the choice of the inverse is for illustrating the suppression
in a more obvious way.
As can be seen as the angular momentum increases the suppression becomes
much bigger, especially at low energies.

The next step is to quantify this effect by means of a more concrete argument.
For instance we can expand Eq.~(\ref{eq:V-central}) at low momenta.
For $p r \ll \sqrt{l+1/2}$ we can expand $j_l$ as
\begin{eqnarray}
j_l(p r) = \frac{(p r)^{l}}{(2l + 1)!!}\,\left[ 1 - 
\frac{1}{2}\frac{(p r)^2}{2 l + 3} +  \mathcal{O}\left( (p r)^4 \right)
\right] \, .
\end{eqnarray}
If we take into account that for radii $m_\pi r > 1$ the OPE potential falls off
exponentially, then for momenta $p, p' \ll \sqrt{l+1/2}\,m_{\pi}$:
\begin{eqnarray}
\langle p', l | V | p, l \rangle &=&
\frac{4\pi}{M_N}\,\frac{p^{l} {p'}^{l}}{{(2l+1)!!}^2} \Big[ 
\int r^{2l+2} dr\,M_N \, V(r) \nonumber \\
&-& \frac{1}{2}\,\frac{{p'}^2 + p^2}{2 l +3}\,\int r^{2l+4} dr\,M_N\,V_C(r)
\nonumber \\
&+& \mathcal{O}({p'}^4 , p^4) \Big] \, ,
\end{eqnarray}
where we have plugged the expansion of the spherical Bessel function
into the integral.
As can be seen, the power law piece of the matrix element is in fact
consistent with the behavior of the lowest-order $l$-wave counterterm 
in pionless theory ($Q^{2l}$), in agreement with naive expectations.
In addition, if we take into account the exponential tail of OPE
it is apparent that
\begin{eqnarray}
\int r^{2k} dr\,M_N\,V(r) \sim \frac{1}{\Lambda_{NN}}\frac{1}{m_{\pi}^{2k-2}}
\, ,
\end{eqnarray}
which means that every $p^{k}$ and $p'^{k}$ within the expansion
of the potential is divided by a $m_\pi^{k}$.
That is, we find the expected breakdown scale of the pionless
power counting.

This begs the question about whether the previous argument can be used
to justify the peripheral demotion of OPE, and in principle the answer
is yes: if $\sqrt{l+1/2}\,m_{\pi} \gg M$ the argument applies
over all the range of validity of the pionful theory.
Using $M \sim 0.5-1.0\,{\rm GeV}$ this happens for $l \gg 12-50$,
for which a demotion of OPE similar to the one found
in pionless EFT will begin to show up.
However this is of little value for practical purposes.
Though we can show that a peripheral demotion indeed exists
it is only apparent for extremely high angular momenta,
far beyond where the partial-wave expansion is
truncated in three-body calculations.
This means that we have to resort to a different type of argument
for analyzing the demotion at moderate angular momenta.
We will do this in the next section.

\subsection{Power Counting Suppression}
\label{subsec:counting}

Here we consider the peripheral wave suppression
from the power-counting perspective.
The idea is to find a relationship between the scales of a two-body system
and the angular momentum.
The arguments we present are in principle tailored for the particular case
of the OPE potential in the singlet channels, in which case the potential
is regular and we do not have to take into account the issue of
cut-off dependence.

We begin by writing the OPE potential in the form
\begin{eqnarray}
\langle p, l | V | p',l \rangle =
\frac{4 \pi}{M_N} \, \frac{1}{\Lambda_{NN}} \, 
f_l(\frac{p}{m_{\pi}} , \frac{p'}{m_{\pi}}) \, ,
\label{eq:ope-reduced-form}
\end{eqnarray}
which underlines the scales that enter into the problem:
$\Lambda_{NN}$ is the natural momentum scale of OPE
and we factor out the $4 \pi / M_N$ that always appears
in non-relativistic scattering.
In the formula above $f_l$ is a dimensionless function defined as
\begin{eqnarray}
f_l(\frac{p}{m_{\pi}}, \frac{p'}{m_{\pi}}) &=&
\int dr\,r^2\,j_l(p r) W_{\rm C}(r) j_l(p' r) \, ,
\end{eqnarray}
where $W_{\rm C}$ -- the central piece of the OPE potential after removing
the $1 / (M_N \Lambda_{NN})$ factor -- is given by
\begin{eqnarray}
W_{\rm C}(r) =
\frac{\sigma \tau}{3}\,m_{\pi}^3\,\frac{e^{-m_{\pi} r}}{m_{\pi} r} \, .
\end{eqnarray}

The scale $\Lambda_{NN}$ determines the strength of the OPE potential
in the singlets, modulo the spin and isospin factors which we have
included in the dimensionless function $f_l$.
How to count $\Lambda_{NN}$ determines whether OPE is perturbative or not.
If we iterate the OPE potential, we can argue on dimensional grounds that
\begin{eqnarray}
\langle V_l G_0 V_l \rangle &\sim& 
\left( \frac{4 \pi}{M_N \Lambda_{NN}} \right)
\times
\left( \frac{M_N Q}{4 \pi} \right)
\times
\left ( \frac{4 \pi}{M_N \Lambda_{NN}} \right) \nonumber \\
&\sim&
\left( \frac{4 \pi}{M_N \Lambda_{NN}} \right) \times 
\left( \frac{Q}{\Lambda_{NN}} \right),
\label{eq:loop-size}
\end{eqnarray}
where $V_l$ stands for the $l$-wave projection of the OPE potential and
$Q$ is either the external momentum $k$ or the pion mass $m_{\pi}$.
From this estimation, we can see that the decision of iterating OPE or not
depends on the dimensionless ratio $Q / \Lambda_{NN}$.

At this point we have the choices $\Lambda_{NN} \sim Q$ (Weinberg)
and $\Lambda_{NN} \sim M$ (Kaplan-Savage-Wise, or KSW).
The first requires the iteration of OPE at LO and the second not.
The numerical value of $\Lambda_{NN} \sim 300\,{\rm MeV}$ lies
in between of what one could consider a soft and a hard scale.
As a matter of fact, none of the previous counting conventions works
for all partial waves: on one hand we have the $^3S_1$ and $^3P_0$
triplets where OPE is thought to be non-perturbative, while on the other
we have the peripheral singlets where the explicit calculations of
Section~\ref{sec:perturbative} show that OPE is perturbative
and probably demoted even with respect to the $\Lambda_{NN} \sim M$
scenario.

The previous mismatch between scaling expectations and numerical calculations
lies in the dimensionless functions and numerical factors
conforming the potential.
Usually we naively assume that these dimensionless factors are of
$\mathcal{O}(1)$ and do not affect the counting.
But if that were the case OPE would be either perturbative or non-perturbative
in all partial waves.
If we take into account these factors, we can have a clearer idea of
what {\it soft} and {\it hard} means for $\Lambda_{NN}$.

To be more specific, let us consider the $l$-wave $T$-matrix of
the OPE potential in the absence of contact-range physics.
If we define the on-shell $T$-matrix as:
\begin{eqnarray}
T_l(k) &=& \langle k,l | T | k,l \rangle \nonumber \\
&=& \langle k,l | V | k,l \rangle +
\langle k, l | V G_0 V | k, l \rangle + \dots \, ,
\end{eqnarray}
then -- according to the naive analysis of Eq.~(\ref{eq:loop-size}) --
we should have the expansion
\begin{eqnarray}
T_l(k) &=& \frac{4 \pi}{M_N \Lambda_{NN}} \sum_{n}^{\infty}
t^{(l)}_{n}(\frac{k}{m_{\pi}}) \, {\left( \frac{Q}{\Lambda_{NN}} \right)}^{n}
\, ,
\end{eqnarray}
where $Q$ is either $k$ or $m_{\pi}$ and $n$ refers to the number of loops.
We expect the coefficients/functions $t^{(l)}_{n}$ to be of
$\mathcal{O}(1)$. On dimensional grounds the $t^{(l)}_{n}(x)$
are functions of $k / m_{\pi}$, as can be deduced
from Eq.~(\ref{eq:ope-reduced-form}).
The dependence on the angular momentum quantum number $l$ enters
via the specific value of the coefficients.
As a consequence, provided that the naturalness hypothesis for $t^{(l)}_n$
is correct, the convergence radius of the series is independent of $l$.

On the contrary, if the convergence of perturbative OPE depends
on the partial wave, then the form of the loop expansion
must take a different form
\begin{eqnarray}
T_l(k) &=& \frac{4 \pi}{M_N \Lambda_{NN}} \sum_{n}^{\infty}
{t'}^{(l)}_{n} (\frac{k}{m_{\pi}}) \,
{\left( \frac{Q}{b_l \Lambda_{NN}} \right)}^{n} \, ,
\end{eqnarray}
where, as in the previous case, $n$ refers to the number of loops and
the coefficients ${t'}^{(l)}_{n}$ are $\mathcal{O}(1)$.
However we include now a factor $b_l$ to account for the different
expansion parameter and convergence radius in each partial wave $l$.
This factor means that $Q / \Lambda_{NN}$ is not the ratio to check
when discussing the power counting of OPE, but rather
$Q / (b_l\,\Lambda_{NN})$.

The only problem left is to determine the expansion parameter
of the power series for $T_l(k)$.
From complex analysis we know that the radius of convergence of this series
is given by the pole of $T_l(k)$ that is closest to the threshold.
However the OPE potential is relatively weak, which means that the poles of
the $T$-matrix will be far from threshold and not easy to find.
With this in mind we can resort to a trick to circumvent this difficulty.
Instead of finding the poles of $T_l(k)$ for the physical value
of $\Lambda_{NN}$, we can change the scale $\Lambda_{NN}$
up to the point that there is a pole at $k=0$,
i.e. a bound state at threshold.

We will give the name $\Lambda_{NN}^*(l)$ to the critical value of
$\Lambda_{NN}$ that generates a bound state at threshold in the $l$-wave.
For $\Lambda_{NN}^*(l)$ the perturbative series at $k=0$ (the location of
the threshold bound state) diverges:
\begin{eqnarray}
T_l^*(0) &=& \frac{4 \pi}{M_N \Lambda_{NN}^*(l)} \sum_{n}^{\infty}
{t'}^{(l)}_{n} (0) \,
{\left( \frac{Q}{b_l \Lambda_{NN}^*(l)} \right)}^{n} \nonumber \\
&\to& \infty \, ,
\end{eqnarray}
where the notation $T_l^*$ indicates that this is the $T$-matrix
for $\Lambda_{NN}^*(l)$.
This means that the expansion parameter is one for the previous series
\begin{eqnarray}
\frac{Q}{b_l \Lambda_{NN}^*(l)} = 1 \, .
\end{eqnarray}
Consequently, for the physical $\Lambda_{NN}$ we have at $k = 0$:
\begin{eqnarray}
T_l(0) &=& \frac{4 \pi}{M_N \Lambda_{NN}} \sum_{n,m}^{\infty}
{t'}^{(l)}_{n} (0) \,
{\left( \frac{\Lambda_{NN}^*(l)}{\Lambda_{NN}} \right)}^{n} \, ,
\end{eqnarray}
from which it is clear that the expansion parameter is 
simply $\Lambda_{NN}^*(l) / \Lambda_{NN}$.

For finite momenta $k \neq 0$ the analysis can be extended easily,
though the conclusions are not as clear-cut as for $k = 0$.
We begin by disentangling the $Q = m_{\pi}, k$ power series explicitly
in the perturbative expansion of the $T$-matrix
\begin{eqnarray}
T_l(k) &=& \frac{4 \pi}{M_N \Lambda_{NN}} \sum_{n}^{\infty}
{t'}^{(l)}_{n} (\frac{k}{m_{\pi}}) \,
{\left( \frac{Q}{b_l \Lambda_{NN}} \right)}^{n} \nonumber \\
&=&
\frac{4 \pi}{M_N \Lambda_{NN}} \sum_{n}^{\infty}
{t'}^{(l)}_{n} (\frac{k}{m_{\pi}}) \, \sum_{r+s = n}
c^{(l)}_{r, s}\,\frac{m_{\pi}^r \, k^s}{b_l \Lambda_{NN}^n}
\end{eqnarray}
where $c^{(l)}_{r,s}$ are coefficients that differentiate the contributions
coming from powers of $m_{\pi}$ and $k$.
The point is that we can rearrange the previous expansion as
\begin{eqnarray}
T_l(k) &=& \frac{4 \pi}{M_N \Lambda_{NN}} \sum_{m}^{\infty}
{t''}^{(l)}_{n} (\frac{k}{m_{\pi}}) \,
{\left( \frac{\Lambda_{NN}^*(l)}{\Lambda_{NN}} \right)}^{n} \, ,
\end{eqnarray}
where the new coefficients ${t''}^{(l)}_{n}$ are defined as
\begin{eqnarray}
{t''}^{(l)}_{n} (\frac{k}{m_{\pi}}) &=& {t'}^{(l)}_{n} (\frac{k}{m_{\pi}}) \,
\sum_{r+s=n} c_{r, s}^{(l)} \left( \frac{k}{m_{\pi}} \right)^s \, .
\end{eqnarray}
If the coefficients $c^{(l)}_{r,s}$ are of order one, the expansion parameter
of the perturbative series will still be
$\Lambda_{NN}^*(l) / \Lambda_{NN}$ for $k \leq m_{\pi}$.
Meanwhile for $k > m_{\pi}$ the expansion parameter is of the order
$k/m_{\pi} \times \Lambda_{NN}^*(l) / \Lambda_{NN}$,
which amount to assuming that $m_{\pi}$ and $k$ play
the same role in the expansion.
However we cannot discard the possibility of relative numerical factors
between the expansions in powers of $m_{\pi}$ and $k$.
For instance, it could happen that the $c^{(l)}_{r,s} \sim {1}/{2^s}$
or that $c^{(l)}_{r,s} \sim {2^s}$, giving a different convergence radius
in terms of $k$ than in terms of $m_{\pi}$.
Though this makes no difference at the conceptual level, this effect
could have a moderate impact when estimating the peripheral demotion
of OPE.
We will briefly discuss this at the end of the section, but we can advance
that the impact is going to be negligible.
Part of the reason lies in the fact that the OPE for $k \geq m_{\pi}$
is very similar to the Coulomb potential, which happens to be
always perturbative except for low energies.
In terms of the perturbative expansion this translates into
the coefficients $c^{(l)}_{r,s}$ having a behavior with respect to $s$ 
that is extremely suppressed (e.g. $1/s!$).

For quantizing the power counting demotion, we compare the expansion
parameter of perturbative OPE with the expansion parameter
of nuclear EFT,
\begin{eqnarray}
\frac{Q}{b_l \Lambda_{NN}} =
\frac{\Lambda_{NN}^*(l)}{\Lambda_{NN}} =
{\left( \frac{Q}{M} \right)}^{\nu_{\rm OPE}(l)} \, .
\end{eqnarray}
Putting the pieces together the actual order of OPE in the $l$-wave singlet
is ${\rm N^{\nu_{\rm OPE}(l)}LO}$, instead of ${\rm LO}$ as in Weinberg or
${\rm NLO}$ as in KSW countings.
It is important to mention here that the existence of a peripheral demotion
$\nu_{\rm OPE}(l)$ hangs on the fact that the scale separation
in nuclear EFT is imperfect.
That is, the clear peripheral demotion we observe depends on the fact
that the separation of scales in nuclear EFT is not particularly good.
If we use the values $Q = m_{\pi}$ and $M \sim 0.5-1.0\,{\rm GeV}$,
the expansion parameter lies theoretically between $1/7$ and $1/3$.
However, we mention in passing that concrete calculations~\cite{Valderrama:2009ei,Valderrama:2011mv,Long:2011qx,Long:2011xw,Long:2012ve}
and analyses~\cite{Birse:2007sx,Birse:2010jr,Ipson:2010ah}
in pionful EFT suggest an expansion parameter closer to $1/3$ than $1/7$.

\begin{figure}[ttt!]
\begin{center}
\epsfig{figure=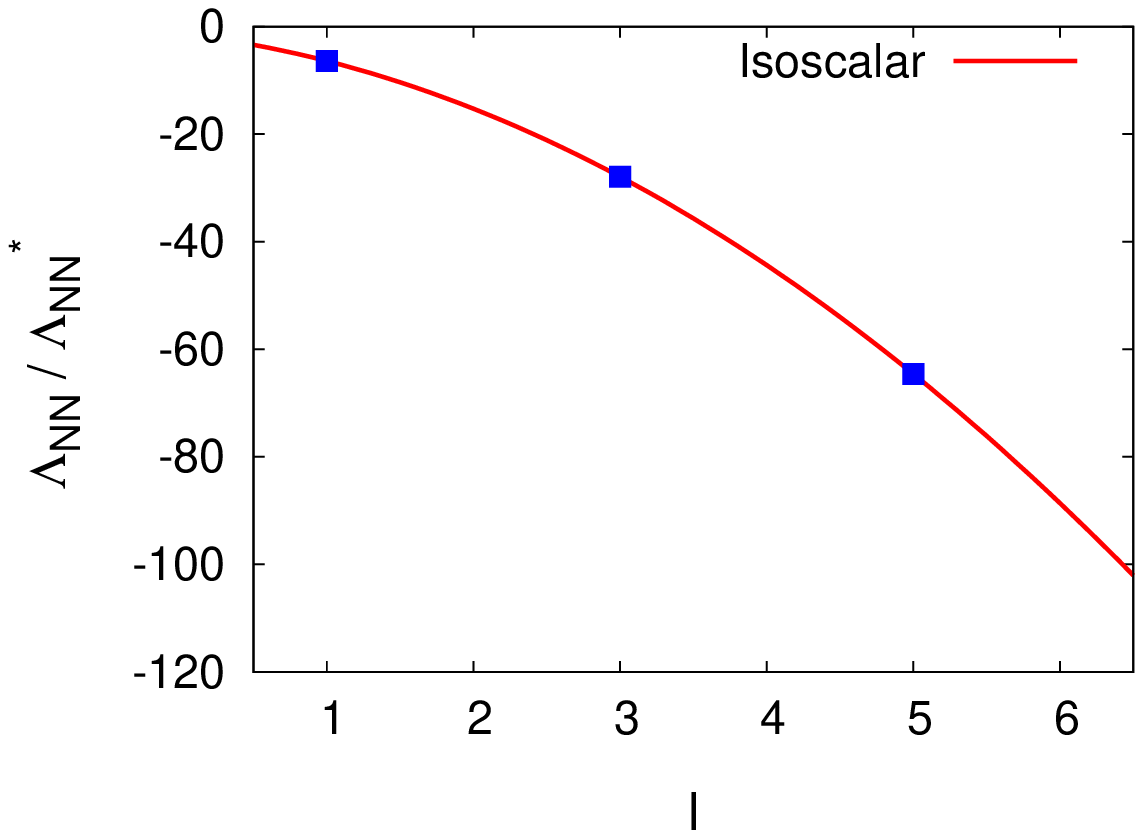,
	height=4.9cm, width=5.4cm}
\epsfig{figure=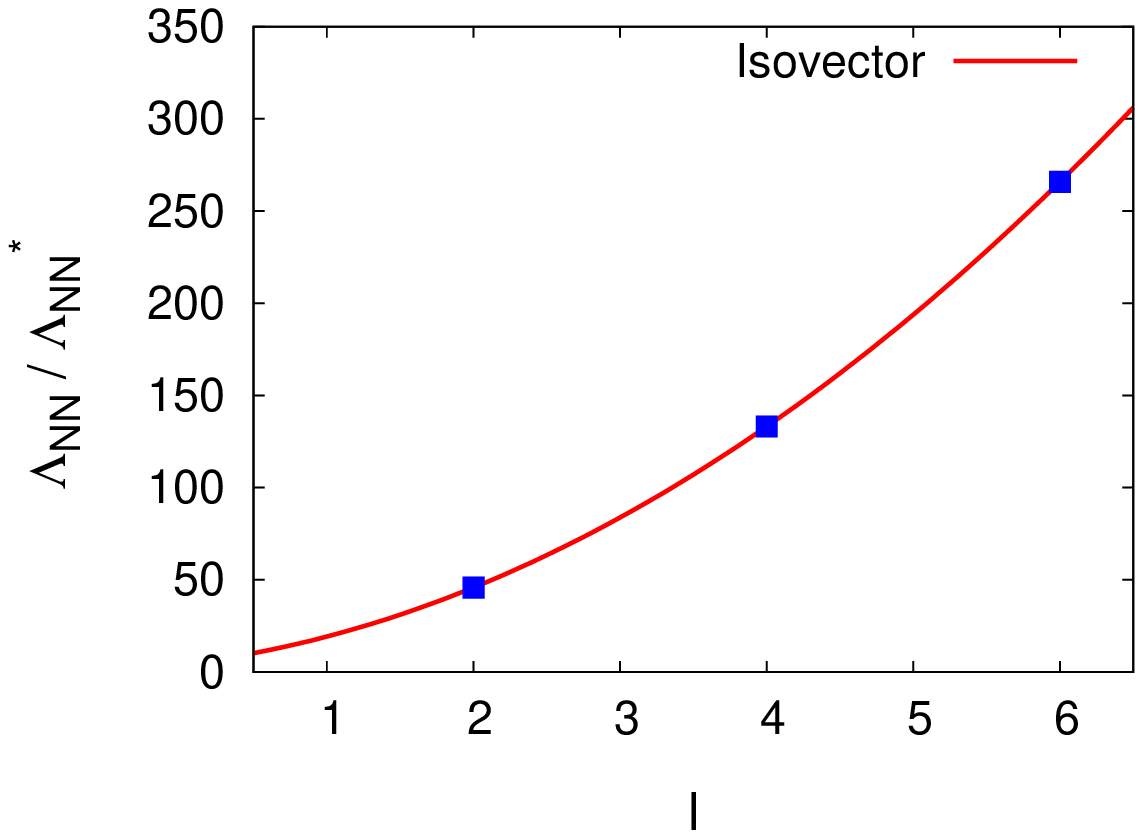,
	height=4.9cm, width=5.4cm}
\end{center}
\caption{Ratio between the actual value of $\Lambda_{NN}$ appearing
in the OPE potential and the critical $\Lambda_{NN}^*$ generating
a bound state at threshold for $l$-wave.
The ratio is calculated for isoscalar and isovector channels. In the isoscalar
case the ratio is negative as the OPE potential is actually repulsive
in the isoscalar channels $^1P_1$, $^1F_3$, $^1H_5$ and so on.
}
\label{fig:rescaling}
\end{figure}

In Fig.~\ref{fig:rescaling} we calculate
the ratio $\Lambda_{NN} / \Lambda_{NN}^*$
for the peripheral singlets as a function of the angular momentum $l$.
For convenience we consider the angular momentum as a continuous
variable~\footnote{That is, we take $l$ real -- instead of integer --
and determine the location of the threshold bound states
from the asymptotic condition $u_l(\infty) = 0$. In practice
we take an infrared cut-off of $R = 40\,{\rm fm}$, though
the results are stable for $R > 10\,{\rm fm}$.}
and include the actual peripheral waves as particular points along this curve.
In the isoscalar partial waves the ratio is negative, the reason being that
the OPE potential is repulsive in isoscalar singlets.
The specific values of the $\Lambda_{NN} / \Lambda_{NN}^*$ ratios
leading to a bound state at threshold are shown in Table \ref{tab:counting},
where we also list the effective order at which OPE enters
in each peripheral singlet.
The power counting is normalized in agreement with the conventions of
the previous discussion: ${\rm LO}$ corresponds to a potential that
has to be iterated to all orders (for instance, the lowest-order contact
interaction in the $^1S0$ partial wave) and ${\rm NLO}$ to the size of
the OPE potential in the KSW counting.
Even for the $^1P_1$ partial wave OPE is slightly demoted with respect to KSW.
Table \ref{tab:counting} only shows partial waves for which OPE is not much
demoted beyond $\rm N^4LO$ (considering the average demotion).
The reason for this is because contributions above this order are unlikely
to enter in any practical EFT calculation in the near future.
The chiral nuclear potential has not been used beyond leading
three-pion exchange (or subsubleading TPE) in full EFT
calculations~\footnote{Though recently the EFT potential has been calculated
one order further and used in first-order perturbation theory for
peripheral nucleon-nucleon scattering, see Ref.~\cite{Entem:2014msa}.}. 
This piece of the potential enters at $\rm N^4LO$ in a Weinberg-inspired
counting~\footnote{This corresponds to $\rm N^3LO$ in the traditional
notation used in Ref.~\cite{Epelbaum:2004fk}, which skips one order
because the $Q/M$ contribution vanishes.} 
and at $\rm N^5LO$ in a KSW-inspired one (for which the convergence is expected
to be slower, leading to a stronger peripheral demotion).
Thus is does not seem necessary to go beyond that point.

The spread in the demotion depends on the expansion parameter taken: for $1/7$
the demotion is relatively mild -- the lowest estimation
in Table \ref{tab:counting} --, while for $1/3$
it is much more obvious.
For instance, in the $^1P_1$ case ${\rm N^{1.0}LO}$ corresponds
to $1/7$ and $\rm N^{1.7}LO$ to $1/3$.
If we take into account the observation that the actual expansion parameter
seems to be closer to $1/3$, then the larger estimations for the demotion
are expected to be more accurate than the lower ones.
However overestimating the demotion can lead to the underestimation
of the theoretical errors in a calculation. In this sense it might be
more cautious to use a value in the middle.

Apart from the uncertainty in $Q/M$ there is a second source of error
in Table \ref{tab:counting}, namely the interplay between
the $k$ and $m_{\pi}$ expansions that we have previously
discussed at the qualitative level.
Addressing this problem formally is beyond the scope of this manuscript
and in fact it has never been done in the literature
for a pionful EFT expansion.
So instead of analyzing in detail the perturbative expansion, we will 
explore the demotion with an alternative definition of $\Lambda_{NN}^*(l)$, 
since this scale is up to a certain extent condition-dependent.
Previously we defined $\Lambda_{NN}^*(l)$ as the $\Lambda_{NN}$
for which there is a bound state at threshold.
As a consequence, the ratio $\Lambda_{NN}^*(l) / \Lambda_{NN}$ corresponds
to the expansion parameter of the $T$-matrix at zero energy.
However, the existence of a low-lying virtual state or resonance also
calls for the iteration of the potential, 
the required strength of the potential being weaker than in the bound state case.
Therefore we release the definition of the scale $\Lambda_{NN}^*(l,k_p)$ as the value of $\Lambda_{NN}$
for which there is a pole at $k = k_p$ (either a bound/virtual state
or a resonance, which means that $k_p$ is a complex value),
this new scale is related to the convergence of the $T$-matrix
for the physical value of $\Lambda_{NN}$ at the location
of the pole
\begin{eqnarray}
T_l(k_p) &=& \frac{4 \pi}{M_N \Lambda_{NN}} \sum_{n}^{\infty}
{t'}^{(l)}_{n} (\frac{k_p}{m_{\pi}}) \,
{\left( \frac{\Lambda_{NN}^*(l,k_p)}{\Lambda_{NN}} \right)}^{n} \, .
\nonumber \\
\end{eqnarray}
In turn we could have used $\Lambda_{NN}^*(l,k_p)$ as the basis of
an alternative definition of the peripheral demotion
\begin{eqnarray}
\frac{\Lambda_{NN}^*(l,k_p)}{\Lambda_{NN}} =
{\left( \frac{Q}{M} \right)}^{\nu_{\rm OPE}'(l)} \, .
\end{eqnarray}
In terms of the scale $\Lambda_{NN}^*$, if we choose a bound state
with $| k_p | = | i \gamma_B | = m_{\pi}$ we will get a smaller
$\Lambda_{NN}^*$ (i.e. more strength for the OPE potential)
and more demotion ($\nu_{\rm OPE}' > \nu_{\rm OPE}$).
On the contrary the virtual state / resonance condition
-- let's say at a momentum $|k_p| = |-i \gamma_V| = | k_R | = m_{\pi}$ --
entails a bigger $\Lambda_{NN}^*$, i.e. a smaller
$\Lambda_{NN} / \Lambda_{NN}^*$ ratio, and 
less peripheral demotion ($\nu_{\rm OPE}' < \nu_{\rm OPE}$).
Actually we are more interested in the possibility that we might have been
overestimating the demotion, which means that we only have to consider
the virtual-state/resonance case.
We have performed the calculations in Appendix \ref{app:resonances}
and checked that the effect of a change in conditions
from a threshold bound state to a resonance is actually tiny
for $\nu_{\rm{OPE}}(l)$, usually of the order of $\Delta \nu_{\rm OPE} \sim -(0.05-0.2)$.
If we compare this change to the uncertainty related to $Q/M$,
which lies in the range $| \Delta \nu_{\rm OPE} | \sim 0.5-2$,
we see that corrections to the threshold bound state
condition can be safely ignored in most partial waves.

\begin{table}[tb]
\caption{\label{tab:counting}
Power counting of OPE for peripheral singlet partial waves.
We show the critical value of $\Lambda_{NN}$ that renders the central potential
non-perturbative in each of the singlets.
The actual power counting assignment for OPE in each of the partial waves
depends on the expansion parameter of nuclear EFT,
which is not known exactly.
However assuming the baseline values of $Q = m_{\pi}$ and
$M \sim 0.5-1.0\,{\rm GeV}$, we expect the expansion parameter
to be between $1/7$ and $1/3$.
Using this range of values, we calculate the OPE demotion
in each partial wave.
}
\begin{ruledtabular}
\begin{tabular}{| c | c | c |}
      $^SL_J$ & $\Lambda_{NN} / \Lambda_{NN}^*(l)$ & ${\rm N^{\nu}LO}$ \\
      \hline
      $^1P_1$ & $-6.40$ & ${\rm N^{1.0-1.7}LO}$ \\
      $^1F_3$ & $-27.9$ & ${\rm N^{1.7-3.0}LO}$ \\
      $^1H_5$ & $-64.6$ & ${\rm N^{2.1-3.8}LO}$ \\
      $^1J_7$ & $-116.4$ & ${\rm N^{2.4-4.3}LO}$ \\
      $^1L_9$ & $-183.3$ & ${\rm N^{2.7-4.7}LO}$ \\
      $^1N_{11}$ & $-265.4$ & ${\rm N^{2.9-5.1}LO}$ \\
      \hline
      $^1D_2$ & $45.8$ & ${\rm N^{2.0-3.5}LO}$ \\
      $^1G_4$ & $133.1$ & ${\rm N^{2.5-4.5}LO}$ \\
      $^1I_6$ & $265.9$ & ${\rm N^{2.9-5.1}LO}$ \\
      $^1K_8$ & $444.0$ &${\rm N^{3.1-5.5}LO}$ \\
      $^1M_{10}$ & $667.4$ &${\rm N^{3.3-5.9}LO}$ \\
   \end{tabular}
\end{ruledtabular}
\end{table}

\subsection{The Peripheral Perturbative Expansion Revisited}

Now that we have a power-counting argument for the centrifugal suppression
of the singlets, we want to check how it stands against concrete calculations.
The approach we find most convenient is the comparison of multiple iterations
of the OPE potential.
We will see how at low energies the expansion parameter of perturbative OPE
is indeed set by $Q /(b_l \Lambda_{NN})$.
The ratio of iterated versus non-iterated diagrams has been already used
in the past as a tool for determining the convergence of
the EFT series~\cite{Fleming:1999ee}, but calculations
have been usually limited to just a few iterations of OPE.
While this might not be a drawback in $S$-wave scattering, peripheral waves
require the evaluation of higher orders of perturbation theory
in order to be able to estimate the expansion parameter.

We will begin with the standard ratio of once iterated OPE versus tree level
for the $l$-wave, which we define as
\begin{eqnarray}
R_l(k) =
\frac{\langle k,l | V G_0 V | k,l \rangle}
{\langle k,l | V | k,l \rangle} \, ,
\end{eqnarray}
where we define the matrix elements above as
\begin{eqnarray}
\langle k,l | V | k,l \rangle &=&
\int_{r_c}^{\infty} dr\,u_k^{[0]}(r)\,V(r)\,u_k^{[0]}(r) \, , \nonumber \\ 
\label{eq:V1} \\
\langle k,l | V G_0 V | k,l \rangle &=&
\int_{r_c}^{\infty} dr\,u_k^{[0]}(r)\,V(r)\,u_k^{[1]}(r) \, , \nonumber \\
\label{eq:V2}
\end{eqnarray}
which roughly corresponds to the diagram ratio already
evaluated in Ref.~\cite{Fleming:1999ee} for the $S$ waves.
The ratio $R_l$ also coincides with the ratio of the one-loop over
the tree level phase shifts
\begin{eqnarray}
R_l(k) = \frac{\delta^{[2]}_l(k)}{\delta^{[1]}_l(k)} \, .
\end{eqnarray}
Once we consider the peripheral demotion factor $b_l$,
we expect $R_l$ to behave as
\begin{eqnarray}
R_l(k) \sim \frac{Q}{b_l \, \Lambda_{NN}} \, . \label{eq:Rl-factor}
\end{eqnarray}
However there is a catch: the ratio $R_l$ can actually be computed analytically
for the OPE potential at low energies and it turns out to be much smaller
than expected from the previous argument.
For showing this we write the perturbative expansion of
the reduced wave functions in terms of a Green's function
\begin{eqnarray}
u_k^{[n+1]}(r) = M_N \int_0^{\infty} d r' G_l(r, r') V(r') \, u^{[n]}(r')
\nonumber \\
\end{eqnarray}
where the Green's function is defined by
\begin{eqnarray}
k\,G_l(r,r') &=&\hat{\jmath}_l (k r) \, \hat{y}_l (k r') \theta(r - r') \nonumber \\
&+& \hat{\jmath}_l (k r') \, \hat{y}_l (k r) \theta(r' - r) \, ,
\end{eqnarray}
with $\hat{\jmath}_l$ and $\hat{y}_l$ the reduced spherical Bessel functions we already
introduced in Section~\ref{sec:perturbative}.
If we take into account that $u_k^{[0]}(r) = \hat{\jmath}_l (kr)$
and expand the reduced Bessel functions in powers of $k r$,
we can evaluate $R_l$ yielding
\begin{eqnarray}
R_l &=& - \frac{\sigma \tau}{2l + 1}
\left( \frac{m_{\pi}}{2^{2 l+1}\,\Lambda_{NN}} \right) \nonumber \\
&\times&
\Big[ 1 + \frac{8 + (l+1) (2l+1)(6l+5)}{2 (2l-1) (2l+3)}\frac{k^2}{m_{\pi}^2} 
\nonumber \\ &+&
\frac{(l+1)(8 + (l+1) (2l+1)(6l+5))}{(2l-1) (2l+3)}\frac{k^4}{m_{\pi}^4} + \dots
\Big] \nonumber \\
&\sim& \frac{Q}{2^{2 l+1}\,\Lambda_{NN}} \, ,
\end{eqnarray}
that is, much more suppressed than expected
(owing to $| \Lambda_{NN} | < 2^{2l+1} | \Lambda_{NN}^* |$).

Does $R_l$ mean that we have underestimated the peripheral demotion?
Not necessarily: the ratio $R_l$, and in general lower-order perturbation
theory, exaggerates the effect of the centrifugal barrier.
The reasons for this mismatch are not completely clear.
It might be related to the interplay of the regular ($\sim r^{l+1}$)
and irregular ($1/r^l$) components of the wave function $u_k$ 
at low energies ($k \ll m_{\pi}$) and intermediate distances
($m_{\pi} r \sim 1$), with the irregular piece giving a larger contribution
and only appearing at higher order perturbation theory.
Be it as it may, the bottomline is that we have to go to higher-order
perturbation theory to probe the expansion parameter.

For exploring higher-order perturbation theory, 
we define the matrix element of the $n$-iterated OPE potential as
\begin{eqnarray}
\langle V_l^{[n]} \rangle &=&
\langle k,l | \underbrace{V G_0 \dots G_0 V}_{\mbox{$n$ insertions of OPE}}
| k,l \rangle \nonumber \\
&=& \int_{r_c}^{\infty} dr\,u_k^{[0]}(r)\,V(r)\,u_k^{[n-1]}(r) \, .
\end{eqnarray} 
We can now define a generalized $R_l^{[n]}$:
\begin{eqnarray}
R_l^{[n]} = \frac{\langle V_l^{[n]} \rangle}{\langle V_l^{[n-1]} \rangle} \, .
\end{eqnarray}
We expect the scaling
\begin{eqnarray}
R_l^{[n]} \sim \frac{Q}{b_l \, \Lambda_{NN}} \, ,
\end{eqnarray}
for large $n$. This expectation is robust and well-grounded as a consequence
of the potential only having a specific fraction of the strength necessary
to bind.

\begin{figure*}[ttt!]
\begin{center}
\epsfig{figure=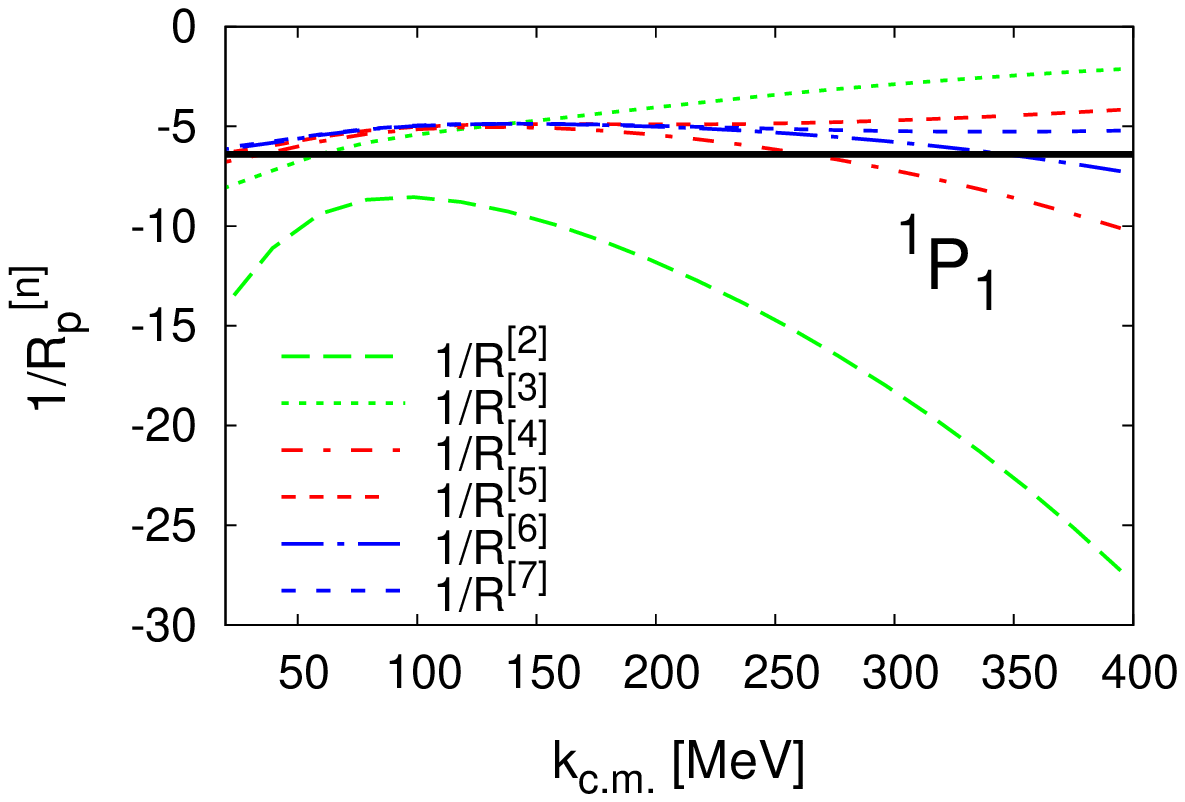,
	height=5.5cm, width=6.5cm}
\epsfig{figure=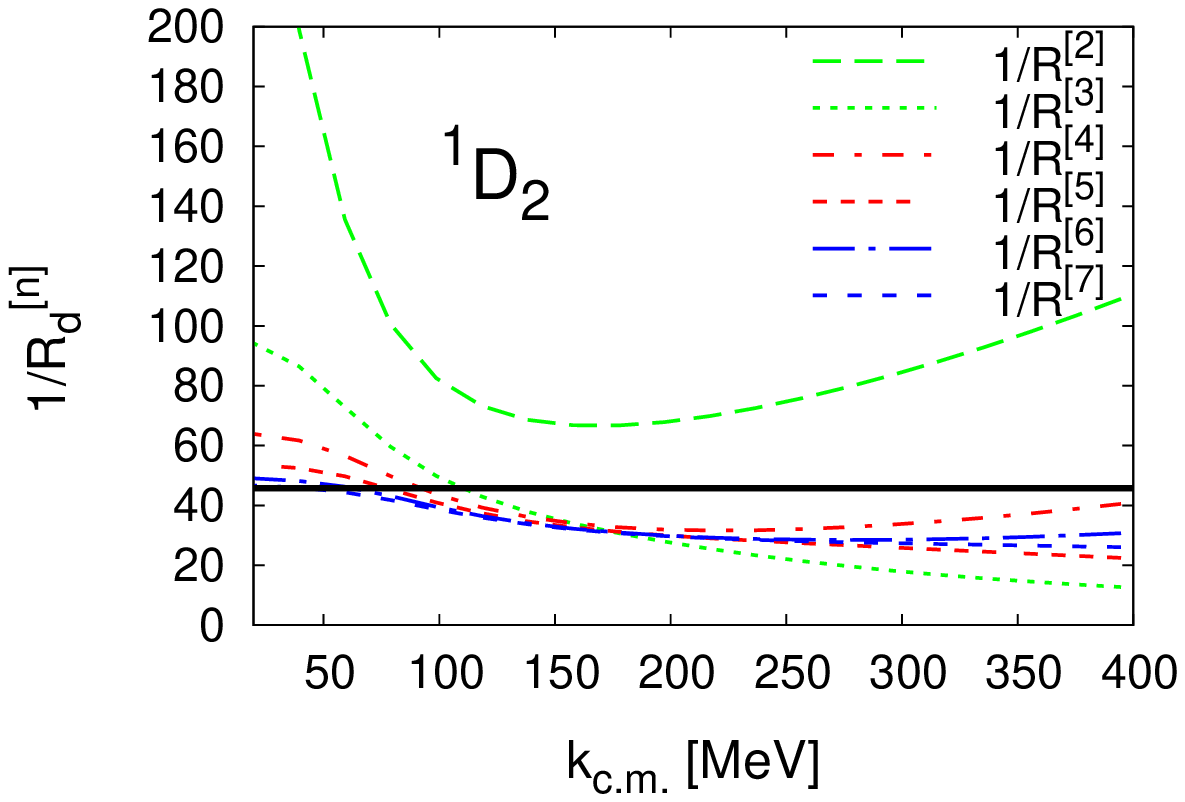,
	height=5.5cm, width=6.5cm}
\epsfig{figure=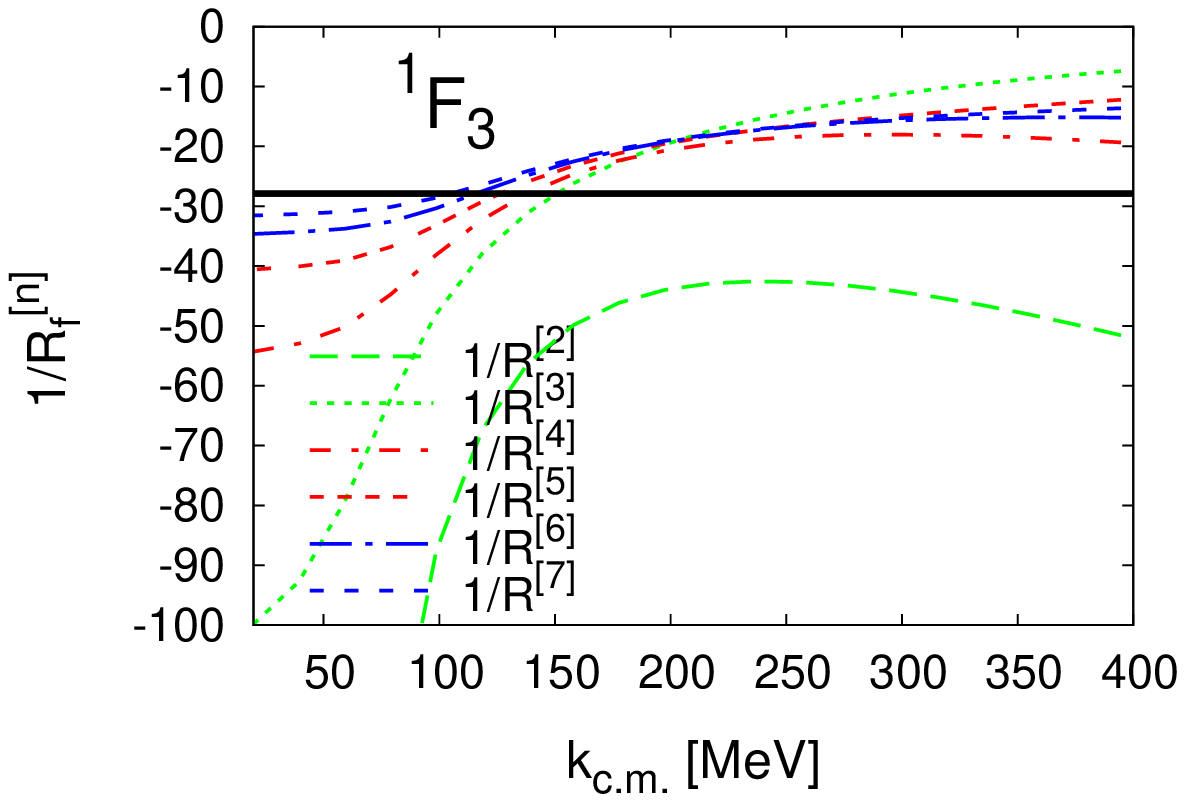,
	height=5.5cm, width=6.5cm}
\epsfig{figure=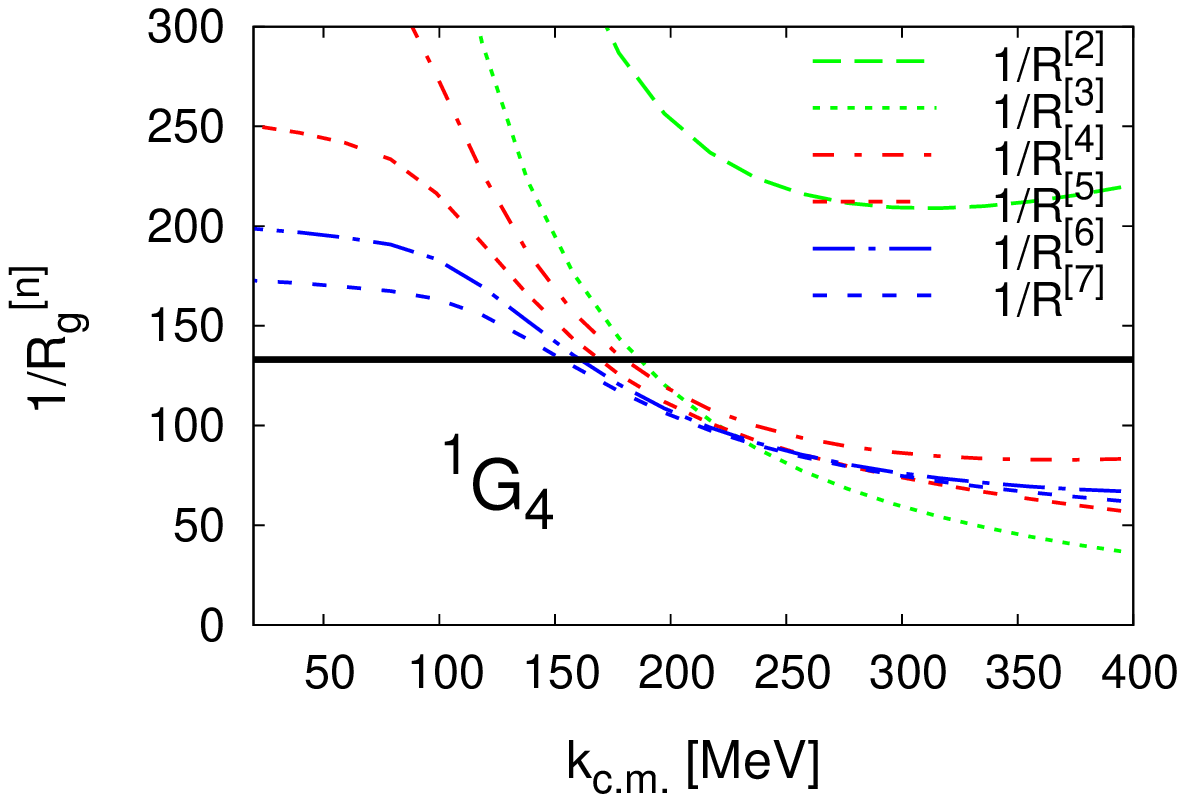,
	height=5.5cm, width=6.5cm}
\epsfig{figure=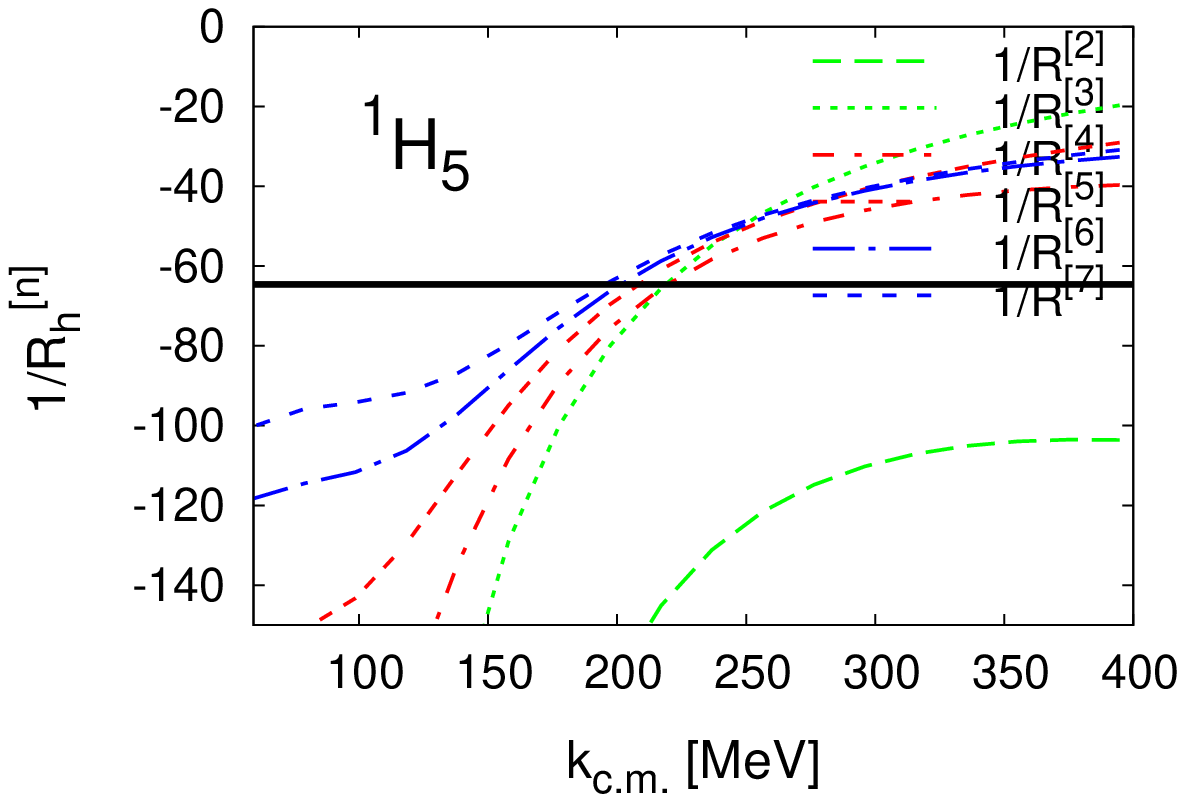,
	height=5.5cm, width=6.5cm}
\epsfig{figure=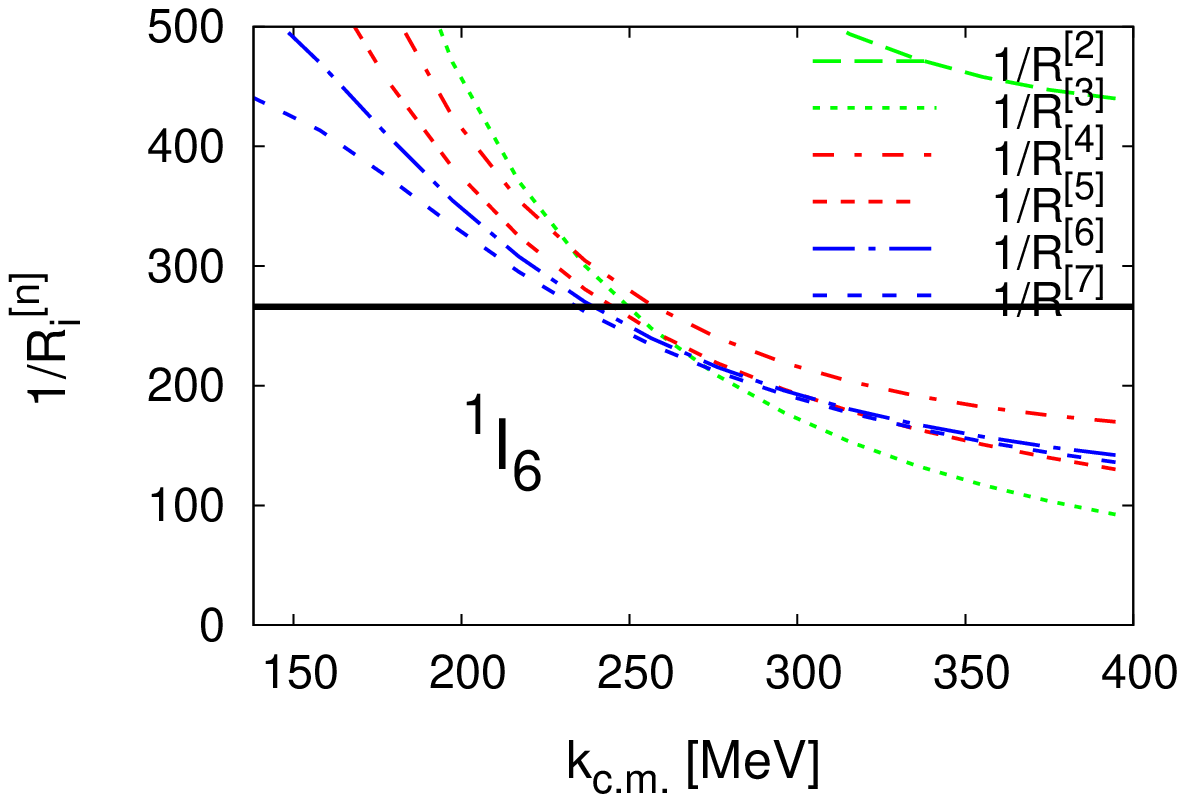,
	height=5.5cm, width=6.5cm}
\end{center}
\caption{Inverse ratios between the $n$-th order over the $(n-1)$-th order
perturbation theory for OPE and the singlet partial waves with $l \leq 6$
(See the main text for a detailed explanation of the ratios.)
The theoretical value to which the ratios should converge
is shown in the black solid line.
}
\label{fig:ratios}
\end{figure*}

We present the calculations of the ratios of the $(n+1)$-th versus the $n$-th
order of perturbation theory in Fig. \ref{fig:ratios}.
For convenience we have chosen the inverse of the ratio $R_l^{[n]}$.
The reason is the inverse of the expansion parameter is a more natural
indication of how well the perturbative expansion converges: the bigger
$| 1/R_l^{[n]} |$ the more convergent.
As a matter of fact, for $k = 0$ we have
\begin{eqnarray}
\frac{1}{R_l^{[n]}} &\sim& \frac{\Lambda_{NN}}{\Lambda_{NN}^*(l)} \, ,
\end{eqnarray}
which when evaluated yield the values of the second column of
Table \ref{tab:counting}, i.e. $1/R_1^{[n]} \sim -6.40$,
$1/R_2^{[n]} \sim 45.8$, $1/R_3^{[n]} \sim -27.9$
and so on.
As can be seen in Fig. \ref{fig:ratios}, $1/R_l^{[n]}$ does converge to
the predicted value at momenta below the pion mass ($k < m_{\pi}$),
where we have extended the previous calculations to seventh order
in perturbation theory to be able to appreciate
the convergence pattern.
For momenta of the order of the pion mass the expansion parameter is
not so good as for $k < m_{\pi}$  but compatible with the EFT expectations,
where we have to take into account that the $Q / b_l \Lambda_{NN}$ expansion
contains powers of both $m_{\pi} / b_l \Lambda_{NN}$ and
$k / b_l \Lambda_{NN}$.
At larger momenta $k > m_{\pi}$ -- and even as we approach
$k \sim M \sim 0.5\,{\rm GeV} $, the expansion still works rather well.
This might be puzzling from the EFT perspective but has a natural explanation
in terms of the form of the central OPE potential.
For large momenta the exponential decay of central OPE is irrelevant and
we are left with a Coulomb-like potential ($V \sim 1/r$).
Coulomb is perturbative at momenta above the inverse Bohr radius.
In the case of central OPE the equivalent is
$k_B \sim m_{\pi}^2 / \Lambda_{NN}$,
which is actually smaller than $m_{\pi}$.
Yet this is a particular feature of central OPE that is not expected to happen
for other contributions of the EFT nuclear potential.

In the calculations of Fig. \ref{fig:ratios} we have reached seventh-order
perturbation theory for checking the expansion parameter.
Yet this is numerically demanding -- more so for low momenta and high partial
waves -- and we limit the analysis to $k_{\rm cm} \geq 20\,{\rm MeV}$
for $l \leq 4$, to $k_{\rm cm} \geq 60\,{\rm MeV}$ for $l = 5$ and
to $k_{\rm cm} \geq 140\,{\rm MeV}$ for $l = 6$.
Though this is a limitation, the calculations are still very meaningful.
For $l \leq 4$ the ratios do converge to the expansion parameter predicted
in Table \ref{tab:counting} with an unrelated method,
providing a cross-check of the calculation.
For $l = 5, 6$ the situation is less clear -- we are limited to higher
momenta and apparently seventh order perturbation theory is not enough
to stabilize the ratios -- but nonetheless the ratios still seem to
converge to the predicted value.
One interesting feature that is evident from the calculations of
Fig. \ref{fig:ratios} is that the low orders of perturbation theory
converges actually faster than the high orders.
The practical implication of this phenomenon is that results at tree level
are more accurate than what is to be expected from the expansion parameter
of the series.
This might in turn point out towards choosing
the higher-order estimates for the demotion.

\subsection{Beyond One-Pion Exchange}

At this point a question arises: how do we extend these ideas to TPE?
NDA predicts leading TPE to be less important than OPE by $Q^2$.
That is, if OPE is ${\rm LO}$ leading TPE is ${\rm N^2LO}$~\footnote{
Here we follow consistency instead of convention: traditionally
leading ${\rm TPE}$ is labeled as ${\rm NLO}$ because 
in Weinberg's counting there is
no contribution to the nuclear potential that is $Q$-times
smaller than OPE. 
If we are accounting for peripheral suppression, which in general
are not integer powers of $Q$, this convention is no longer useful.
}.
But we have seen here that OPE is probably demoted by a larger factor 
for $D$ and higher waves.
It is natural to expect that a similar demotion will happen for TPE.
The size of this effect will naturally depend on the partial wave and
it might very well happen that in exceptional cases TPE might enter
before OPE.
This might explain why there are a few peripheral waves
for which TPE is required to explain the phase shifts deduced from experimental data,
such as $^1D_2$, $^3P_2$ or $^3D_3$~\cite{PavonValderrama:2005uj}.

For OPE we have developed a power counting argument
for the peripheral demotion.
But this idea depends on an especial feature of the OPE potential
in singlet waves: this is a regular potential
that does not require regularization.
For TPE the situation is different, it is a badly divergent potential
at short distances and requires regularization.
Exactly the same happens with OPE in the triplets.
As a consequence the argument we have developed
here cannot be applied either for TPE or
for OPE in the triplet channels.
There are strategies to cope with this, though they will require serious
scrutiny to check whether they work. 
The most obvious one is to renormalize these partial waves: if we add
a contact-range interaction we might be able to apply
the previous argument.
The drawback of this idea is that it mixes short- and long-range physics.
The factor by which we have to rescale TPE for having a bound state
at threshold depends on the scattering volume of the channel
(before rescaling TPE), which fixes the contact-range coupling.
The meaning of this is that the rescaling factor is contaminated
by the physical scattering length, which is undesirable.
It might happen though that the effect is negligible,
as happened with the bound-state versus resonance condition
in the previous discussion.

If we strive for a solution that is manifestly independent of the existence
of short-range physics two possibilities come to mind.
The first is Birse's approach to tensor OPE~\cite{Birse:2005um},
which adapts a series of techniques from molecular physics
to study when tensor OPE is perturbative.
A limitation of this approach is that it is formulated in the chiral limit,
where the range of the OPE potential becomes infinite and it is thus
similar to the potentials found in atomic physics.
A possibility is to study the cut-off at which deeply bound states
happen with TPE.
Deeply bound states are non-physical bound states that occur
when we have attractive singular interactions, such as TPE.
They are of no consequence because their binding momenta are beyond
the range of applicability of EFT. In addition techniques have been
developed to get rid of them~\cite{Nogga:2005hy}.
The point is that the more peripheral the wave the harder the cut-off
for which deeply bound states appear.
This might in turn gives us information
on the quantitative nature of the partial wave suppression.
\section{Conclusion}
\label{sec:conclusion}

In this work we have analyzed from the EFT perspective -- and in the particular
case of the spin-singlet channels -- the common wisdom observation
that pion exchanges are perturbative in peripheral waves.
For this we have studied the convergence of the perturbative expansion of
the phase shifts numerically up to fourth order in perturbation theory.
This calculation -- which has been done for the first time
up to such an order in this work -- indicates that pions
are perturbative in the singlets.
In fact, the multiple iterations of one-pion-exchange potential turn out to be much more suppressed
than expected even in a power counting (Kaplan-Savage-Wise)
in which one-pion-exchange potential is treated as subleading.

To understand this pattern we have made use of a power counting argument
to determine the actual demotion of one-pion-exchange potential with respect to leading order.
The idea is to rescale the strength of the one-pion-exchange potential up to the point
in which a bound state is generated at threshold.
This critical strength can be translated into a critical $\Lambda_{NN}^*$
-- which does not correspond with the physical $\Lambda_{NN}$
but will in general be softer -- for which the perturbative expansion
does not converge.
The ratio $\Lambda_{NN}^* / \Lambda_{NN}$ corresponds
to the expansion parameter of perturbative one-pion-exchange potential,
which turns out to be quickly convergent.
We have checked this prediction against concrete calculations,
confirming the EFT argument.
Actually even the $^1P_1$ partial wave is demoted beyond next-to-leading order
and more peripheral waves can be demoted up to the point of being
less important than subleading two-pion-exchange potential in naive dimensional analysis (it is yet to be seen
how demoted will be leading and subleading two-pion-exchange potential in peripheral
waves).
 
The importance of the peripheral demotion is not merely academic.
It has applications in few-body calculations, where the demotion can be used
to improve and optimize the calculations.
The way in which this is achieved is by only including the necessary number
of iterations in the peripheral waves and by ignoring partial waves where
tree-level one-pion-exchange potential is already higher order than the order of the calculation. 
Actually this is analogous to the common practice of ignoring the partial waves
with angular momentum bigger than a certain critical value
($l \geq 5$ in most applications).
The difference is that here we systematize this practice in a way
that is compatible with the EFT expansion, providing guidelines
for future few-body calculations in nuclear EFT.

Yet this manuscript deals only with the one-pion-exchange potential in the peripheral singlets.
For the peripheral demotion to be useful in few-body calculations,
we have to extend the present analysis to peripheral triplets
and also to the two-pion-exchange potential.
This analysis is underway, though the tools that will be required
are different that the ones we have used here.
The fundamental difference of one-pion-exchange potential in peripheral triplets and of two-pion-exchange potential is that these potentials are singular, diverging faster than $1/r^2$
at distances below the pion Compton wavelength.
Thus the calculation of their peripheral demotion will require
the development of more sophisticated arguments that take into
account the existence of a finite cut-off and how it relates
to the other scales in the problem.

\begin{acknowledgments}
UvK and MPV would like to thank Monique Lassaut
for useful discussions on the convergence of perturbation theory
in the presence of virtual states.
UvK was supported in part by the U.S. Department of Energy, Office of Science, 
Office of Nuclear Physics, under award number DE-FG02-04ER41338, and
by the European Union Research and Innovation program Horizon 2020
under grant agreement no. 654002.
MPV was partially supported by the Fundamental Research Funds for the Central Universities and 
by the National Natural Science Foundation of China (NSFC) through grant numbers 11375024 and 11522539.
BWL was supported in part by the NSFC through grant number 11375120.
\end{acknowledgments}

\appendix

\section{Integral Representation for the Phase Shifts}
\label{app:integral}

In this short appendix we derive Eq.~(\ref{eq:integral}),
the integral representation of the phase shifts
that we later use in perturbative calculations.
The starting point is to consider the reduced Schr\"odinger equations
for the free problem and for two particles interacting via a central
potential $V$:
\begin{eqnarray}
- v_l'' + \frac{l(l+1)}{r^2}\,v_l(r)
&=& k^2 v_l \, , \\
- u_l'' + \left[ 2\mu\,V(r) + \frac{l(l+1)}{r^2} \right]\,u_l(r)
&=& k^2 u_l \, .
\end{eqnarray}
Now we construct a Wronskian identity involving the free and
non-free wave functions.
For this, we first multiply the free equation above by
the non-free wave function $u_l$ and vice versa.
Then, we calculate the difference between the two expressions we obtain.
We end up with
\begin{eqnarray}
(v_l \, u_l'' - v_l'' \, u_l) - 2\mu\,V(r)\,u_l\,v_l = 0 \, .
\end{eqnarray}
The terms in the brackets are the derivative
of $(v_l \, u_l' - v_l' \, u_l)$.
We can thus integrate the expression above to reach:
\begin{eqnarray}
(v_l \, u_l' - v_l' \, u_l){\Big|}^R_{r_c} =
2\mu\,\int_{r_c}^R\,V(r)\,u_l(r; k)\,v_l(r; l)\,dr \, .
\end{eqnarray}
Now we simply take  $v_l =\hat{\jmath}_l(k r)$ and $u_l$
as in Eq.~(\ref{eq:asymptotic}).
After removing the infrared and ultraviolet cut-offs
($R \to \infty$, $r_c \to 0$), we obtain
\begin{eqnarray}
k\,\tan{\delta}_l = -2\mu\,\int_{0}^{\infty}\,V(r)\,u_k(r)\,v_k(r)\,dr ,
\end{eqnarray}
that is, Eq.~(\ref{eq:integral}).

\section{Peripheral Demotion and Resonances}
\label{app:resonances}

In this appendix we discuss the different choices for the definition of
$\Lambda_{NN}^*$ and which effect do they have on the peripheral demotion
of the OPE potential.
As we will see the effect of changing the threshold bound state condition
by a different condition is going to be rather small.
This is so because the choice of rescaling up to the point of having
a bound state at threshold, instead of at a finite binding energy,
is not arbitrary.
The EFT description of a bound state with the OPE potential will involve
the scales $m_{\pi}$, $\Lambda_{NN}^*$ and the wave number
$\gamma = \sqrt{M_N B}$, with $B$ the binding energy.
By requiring $\gamma = 0$ we eliminate one of the variables in the problem:
the only scales left are $m_{\pi}$ and $\Lambda_{NN}^*$, which combine
by means of a numerical factor to give an expansion parameter of one
if there is a threshold bound state.
That is, $\gamma = 0$ is the easiest and most convenient choice to isolate
the scale $\Lambda_{NN}^*$.

On the contrary if $\gamma \neq 0$ the EFT expansion involves two ratios:
$m_{\pi} / \Lambda_{NN}$ and $\gamma / \Lambda_{NN}$.
The failure of the expansion at a particular $\Lambda_{NN}$ for a pole
at $\gamma \neq 0$ does not necessarily mean that $\Lambda_{NN}$
is the $\Lambda_{NN}^*$ we are looking for.
There might be a mismatch -- probably a small one -- between the two
owing to the different numerical factors in each of the subexpansions.
If we were able to take these details into account, it might very well happen
that we end up with the original scale generating the threshold bound state.

Yet we will ignore these complications here.
The point is to check the robustness of the estimations we have made.
For that we will assume a different choice for the location of a pole
in the amplitude and repeat the analysis of Section \ref{subsec:counting}.
Instead of $k_p = i \gamma_B = 0$ for a bound state at threshold,
we will consider the case of a virtual state or resonance: $k_p = - i \gamma_V$
or $k_p = k_R$, where $\gamma_V$ is real and $k_R$ is complex.
This will translate into a new value of the critical $\Lambda_{NN}^*$
that depends on the position of the pole:
$\Lambda_{NN}^* = \Lambda_{NN}^*(l, k_{\rm pole})$
with $| k_{\rm pole} | \sim Q$.
Proceeding as in the $k_{\rm p} = 0$ case, we define the peripheral demotion as
\begin{eqnarray}
\frac{\Lambda_{NN}^*(l, k_{\rm pole})}{\Lambda_{NN}} =
{\left( \frac{Q}{M} \right)}^{\nu'_{\rm OPE}(l, k_{\rm pole})} \, ,
\end{eqnarray}
where we have added the $'$ to distinguish the new estimations
from the previous ones.

Virtual states and resonances are poles in the scattering amplitude
in the second Riemann sheet of the complex energy plane.
They are easy to locate in the case of contact-range potentials.
For a finite-range potential, finding the virtual and resonant states
is technically more challenging, due to the difficulty of choosing
the second Riemann sheet in a numerical calculation.
As we are interested in peripheral waves, the most natural outcome when
we reduce the strength of the potential is that a bound state
eventually becomes a resonance.
This offers a simplification: resonances have a very clear signature
in physical scattering.
There is a jump of $180$ degrees in the phase shifts near the location
of the resonance, where the jump will be very steep if the resonance
is narrow.
In addition at a energy similar to the real part of the resonance pole
the scattering amplitude saturates the unitarity bound and the phase
shift reaches $90$ degrees.
Therefore the criterion we are going to use for $\Lambda_{NN}^*(l, k_R)$ will be
\begin{eqnarray}
\cot\,{\delta(k = m_{\pi})} = 0 \, ,
\end{eqnarray}
which in general will imply $| k_R | > m_{\pi}$, but only by a small amount
if the resonance is narrow.

The change in $\nu_{\rm OPE}(l)$ for the resonance condition are show
in Table \ref{tab:resonances}.
In general the new condition only entails a tiny change in $\nu_{\rm OPE}(l)$
in the $-(0.05-0.2)$ range. This change is an order of magnitude smaller
than the changes in $\nu_{\rm OPE}(l)$ related to the uncertainty
in the EFT expansion parameter $Q/M$ and hence
can be safely ignored.

\begin{table}[ttb]
\caption{\label{tab:resonances}
Changes in the peripheral demotion of OPE if we change the definition of
the critical $\Lambda_{NN}^*(l, k_{p})$.
While the original definition involved the critical value of $\Lambda_{NN}$
that generates a bound state at threshold,
here we use an alternative definition of $\Lambda_{NN}$ that generates
a phase shift of $\pi/2$ at $k = m_{\pi}$.
This definition is in turn related to the existence of a nearby resonance
at $k_R \sim m_{\pi} - i \gamma$ (where the approximation will be better
the smaller $\gamma$).
From the previous and new values of $\Lambda_{NN}^*$ we can estimate the change
in the peripheral demotion with respect to the shallow bound state condition,
shown in the $\Delta \nu = \nu' - \nu$ column, where we have assumed $Q/M$
to change between $1/7$ and $1/3$.
As can be appreciated the changes are tiny and not worth considering.
Owing to the resonances becoming narrower for higher partial waves we could
only compute the shifts $\Delta \nu$ for $l \leq 6$. For higher partial
waves we expect the shifts to be even tinier.
}
\begin{ruledtabular}
\begin{tabular}{| c | c | c | c |}
      $^SL_J$ &
      $\Lambda_{NN} / \Lambda_{NN}^*$ &
      $\Lambda_{NN} / \Lambda_{NN}^*(m_{\pi})$ &
      $\Delta \nu = \nu' - \nu$ \\
      \hline
      $^1P_1$ & $-6.40$ & $-4.86$ &$-(0.14-0.25)$ \\
      $^1F_3$ & $-27.9$ & $-24.9$ &$-(0.06-0.10)$ \\
      $^1H_5$ & $-64.6$ & $-59.1$ &$-(0.05-0.08)$ \\
      \hline
      $^1D_2$ & $45.8$ & $33.31$ & $-(0.16-0.29)$ \\
      $^1G_4$ & $133.1$ & $119.7$ & $-(0.05-0.10)$ \\
      $^1I_6$ & $265.9$ & $250.2$ & $-(0.03-0.05)$ \\
   \end{tabular}
\end{ruledtabular}
\end{table}

\end{document}